# Investigation of non-Hermitian and Hermitian models of Altermagnets

*Partha Goswami*

Email: *physicsgoswami@gmail.com*

**Abstract:** Insulating altermagnets like MnTe exhibit spin configurations where opposing spins are not only aligned antiparallel but also rotated relative to each other. This is an arrangement reminiscent of antiferromagnetism with a twist of spin canting. This study investigates a model Hamiltonian that captures the essential physics of such systems, incorporating key interactions including Dzyaloshinskii-Moriya and conventional exchange terms, relativistic spin-orbit coupling, and d-wave and g-wave orderings. Non-Hermitian dynamics are introduced through complex potentials that simulate energy dissipation and amplification. The paper delves into the behavior of the quantum geometric tensor and the emergence of the quantum anomalous Hall effect within the topologically insulating regime. It also broadens the scope to encompass non-Hermitian metallic altermagnets, focusing on phases characterized by symmetry-breaking d-wave and g-wave order parameters.

## 1.Introduction

In ferromagnetic (FM)materials, net magnetization is accompanied by an energy split between upward and downward spin bands, whereas antiferromagnetic (AFM) materials exhibit degenerate spin-up and spin-down energy bands with identical dispersion in momentum space due to lattice translation or inversion symmetry. In the latter sublattices are related by inversion or translation, resulting in spin-degeneracy of electronic bands. However, the recent exploration of altermagnetism (AM) **[1-21]**, a novel subclass of magnetic materials with broken time-reversal symmetry (TRS), unveils a distinct collinear magnetic order (extended AM includes 'non-collinear' spin states **[14]**) where neighboring spins are antiparallel with zero net magnetization. In AMs, lattice rotation connects opposite-spin sublattices inducing nonrelativistic spin splitting (SS) in momentum space with rotational symmetry-dictated sign patterns unlike AFMs. In other words, AMs' non-relativistic SS originates from rotational and/or reflectional connections between sublattices of opposing spins.  This phenomenon is substantially larger and distinct from relativistic spin-orbit coupling effects in terms of symmetry classification. Thus, AMs exhibit two defining properties **[4-10]** are (a) net magnetization absence, unaffected by broken TRS, and (b) momentum-dependent electronic band spin splitting, independent of spin-orbit coupling. To get these points across in fewer words, AM is a novel magnetic phase that transcends the traditional dichotomy between ferromagnetism (FM) and antiferromagnetism (AFM), exhibiting globally compensated magnetization and directional spin polarization. The exploration of AMs, originally motivated by spintronics research, has unlocked fresh perspectives in the field of broken-symmetry phases exhibiting *d*-wave, or other non-*d*-wave ordering characteristics. These materials underpin promising properties including spin-polarized conductivity, spin-transfer torque, anomalous Hall effect, tunnelling, and giant magnetoresistance, crucial for advancing next-generation memory devices, magnetic detectors, and energy conversion technologies **[13]**.

The emergence of AMs', diverging from established FM and AFM paradigms, has sparked considerable interest of condensed matter physics community specially due to its unconventional vanishing of net magnetization stemming from alternating direct and momentum space magnetic moment ordering, which is in contrast with TRS-breaking ferromagnets. Recent studies [11,15] have confirmed room-temperature antiparallel magnetic ordering in metallic $RuO_2$, solidifying its importance in altermagnetism. The compound $RuO_2$ corresponds to tetragonal crystal/lattice system (Symbol: $P4_2/mnm$, Point Group: 4/mmm). Ongoing research [16] highlights $RuO_2$'s significance, while other rutile-structured materials, like insulating $FeF_2$ [17] and $MnF_2$ [10,18], exhibit characteristic spin-polarization order, breaking TRS. The insulating (metallic) AM model to be probed is schematically illustrated in Figure 1(a) (1(b)). In both the cases, the resulting order exhibits symmetry under combined real space fourfold rotational symmetry $C_4$ and the time-reversal symmetry $\mathcal{R}$. The system in 1(a), however, is also invariant under $\mathcal{R}$ and real-space mirror reflections, with the two mirror planes intersecting the lattice shown as dashed lines. The nearest neighbor and next-nearest neighbor hops are indicated in this figure, with the unit cell marked by a square with solid lines as the sides. Altermagnetism is anticipated to be a pervasive phenomenon in nature, occurring in both three-dimensional and possibly far-from-equilibrium two-dimensional crystals, and encompassing the entire spectrum of conduction and insulation types, from topological insulators and metals to superconductors. In fact, recent research [19] has identified a regime where a two-dimensional crystal can form if the system is driven out of equilibrium. The crystal lattices of AMs generally display parity-time (*PT*) symmetry [20], but can also exhibit broken *PT* symmetry in certain situations [14]. *PT*-symmetric systems, marked by complex potentials, interact with their environment (as we will see in section 2). Energy exchange is dictated by the potential's imaginary component: positive values correspond to energy gain, negative values to energy loss. Importantly, PT symmetry ensures equilibrium, balancing energy acquisition and relinquishment. The *PT* symmetry disruption occurs through three distinct mechanisms: [a] breaking *P* symmetry while preserving *T* symmetry, [b] breaking *T* symmetry while preserving *P* symmetry, and [c] concurrent disruption of both *P* and *T* symmetries. Additionally, as observed in ref.[21], the space-time reflection symmetry is compromised when *P* and *T* are disrupted differently. In AMs, the order parameter symmetries and connection between opposite-spin sublattices are characterized by unique features that distinguish them from conventional magnetic phases. For instance, they have even spin-polarization order in their band structure, namely *d, g,* or *i*-wave order and the opposite-spin sublattices are connected through crystallographic rotation transformations. While *g*-wave order means angular momentum 4, *i*-wave order corresponds to 6. These are distinct features compared to other magnetic phases. These features set them apart from conventional Ferromagnets (FMs) and Antiferromagnets (AFMs).

As mentioned above, the AM materials [1-14] blend ferromagnet-like properties, including robust time-reversal symmetry-breaking responses and spin polarization, with antiferromagnet-like traits, specifically, antiparallel magnetic crystal order and vanishing net magnetization. The materials emerge as a distinct third magnetic phase, reconciling this seeming inconsistency. This perspective requires detailed analysis of the model Hamiltonians related to this novel (*d*-wave and higher even-parity wave) magnetic phase. We present our investigation results in this direction through this communication. In fact, our investigation concentrates on materials that violate $\mathcal{R}$ symmetry and $C_4$ [22,23], while preserving their combined symmetry, $C_4\,\mathcal{R}$. To provide additional context, we commence with a non-Hermitian two-dimensional, spinless toy model on a square lattice with lattice constant $a$, comprising two sublattices A and B, which will be discussed in section 2 in details. A symmetry analysis reveals that the Hermitian part of our Hamiltonian, breaks TRS ($\mathcal{R}$) and four-fold rotational symmetry ($C_4$) along the z-axis, yet maintains their collective symmetry $C_4\,\mathcal{R}$. In this context, each lattice site is envisioned as hosting a hypothetical $p$ orbital, as schematically illustrated in Figure 1(a). This conceptualization aims to lower the local symmetries of the A and B sites from $C_4$ to $C_2$, thereby preventing the lattice from being reduced to a simple cubic structure with one atom in a unit cell. In addition, we include an imaginary part in our model which corresponds to the non-Hermitian dissipation/gain. Subsequently, we explore a model Hamiltonian for MnTe, an insulating AM (IAM), involving the d-wave and the g-wave pairing order, where the sublattice and the spin degrees of freedom will be given due consideration. The plots of the energy eigenvalues of the model Hamiltonian exhibit spectral gaps (see Figure 2) at the high symmetry points $\Gamma$ (0,0), $M$ ($\pi$, $\pi$), X($\pi$, 0) and Y(0, $\pi$)) of the system. This exercise is followed by the presentation of models of the $\nu$ -wave ($\nu$ = $d$-wave/$g$-wave) ordered metallic altermagnets (MAM) on a square lattice with spin degrees of freedom only. The Hamiltonian of the MAM includes kinetic energy term, even-parity wave order parameters, Rashba spin-orbit coupling (RSOC) strength, and a term which represents the momentum-independent magnetization component perpendicular to the plane of the 2D system. The magnetization component will arise from either the ferromagnetic insulator substrate or an external perpendicular magnetic field. Additionally, there would be inclusion of a term $\gamma$ for intrinsic loss to investigate the non-Hermiticity aspect. A symmetry analysis reveals that, when magnetization is zero, the Hermitian part of our Hamiltonian maintains their collective symmetry $C_4\,\mathcal{R}$. It is worth noting that RSOC violates inversion symmetry without undermining this combined symmetry. The $C_4\mathcal{R}$ symmetry is a distinguishing feature of AMs [4,5,19,22-24], which inherently exhibit spin-polarized bands. Finally, in section 2, we explore a model Hamiltonian, involving the d-wave and the g-wave pairing order, where the spin and the sublattice degrees of freedom will be given due consideration. The inducement of the Kramers degeneracies at the high symmetry points has been reported earlier[25]. Our investigation of spinful fermions on a square lattice with two sublattices shows that, whereas Kramers degeneracies occur at X($\pi$, 0), Y(0, $\pi$), and $M$ ($\pi$, $\pi$) points for the $d$-wave ordered $d$AM, for the $g$-wave ordered system the degeneracies occur at $\Gamma$(0, 0), and $X(\pm\pi$, 0). Despite the $C_4\mathcal{R}$ symmetry being disrupted by non-zero magnetization, the overall system retains the $C_2$ rotational symmetry. Due to the presence of this crystalline symmetry $C_{2z}$, the system allows the determination of weak topological indices based on the eigenvalues of the $C_{2z}$ operator at $C_{2z}$ invariant momenta. One needs to follow the principles outlined in the Fu-Kane-Mele formula [26-29] for this task. Furthermore, the IAMs, exhibiting a nonzero Chern number, display topologically nontrivial features such as the anomalous Nernst effect (ANE) and the quantized anomalous Hall conductance (QAHC)[24] (the Chern number is quantized for IAMs only), even in the absence of conventional magnetic ordering. In section 3 of this paper we will calculate QHC. A relevant question is why consider a 2D model for theory of altermagnetism when compounds like $RuO_2$ has a 3D crystalline structure. The rationale lies in the fact that by employing a 2D model approximation, one can focus on the key aspects of the system's behavior, such as magnetic ordering or symmetry breaking, without getting bogged down in the full 3D structure. This is especially useful when the 2D limit of a thin film can still capture the relevant physical phenomena of interest. Additionally, the complexity of the full 3D Hamiltonian can be difficult to deal with analytically or computationally. Generally speaking, in many cases, the low-energy physics of a material can be approximated by a 2D model, particularly when interactions, magnetic properties, or phenomena like frustration or anisotropy are primarily confined to a plane.

The study of open systems [30-32], featuring inherently non-Hermitian Hamiltonians due to interactions with external baths, has sparked considerable interest in exceptional points (EPs) [30-32] recently, prompting rapid growth in non-Hermitian topological systems research. In fact, non-Hermitian systems have attracted significant interest in recent years [33-38], driven by their unique characteristics and benefits over conventional Hermitian systems, particularly in terms of spectral properties and topological features, which prove valuable in various domains, including optical and photonic systems, quantum computing, non-reciprocal light propagation, and precision measurements. Specifically, non-Hermitian topological systems are a focal point of research in condensed matter physics. An important aspect of non-Hermiticity of a Hamiltonian is that it exhibits exceptional points (EPs), at specific momenta, where the eigenvalues of the non-Hermitian Hamiltonian become degenerate, and the corresponding eigenvectors also coalesce [33–38]. One needs to visualize the spectrum of the Hamiltonian in terms of a complex plane, and exceptional points appear as singularities or branch-point singularities within non- Hermitian eigenvalue manifolds. Leveraging EPs in non-Hermitian systems has emerged as a fruitful area of research. Notably, near EPs, degenerate modes exhibit sub-linear splitting in response to minor perturbations, enabling enhanced measurement sensitivity and informing the design of EP-enhanced sensors.

The Quantum Geometric Tensor (QGT) [39-44] is a specialized mathematical tool, engineered to merge geometric and quantum concepts, with a focus on quantum information theory [38] and condensed matter physics [42]. QGT plays a vital role in characterizing geometric and topological properties of quantum states, defined within the framework of quantum state manifolds, combining the Fubini-Study metric (FSM), which captures geometric properties of quantum pure states, and the Berry curvature (BC), elucidating phase and geometric effects arising from parameter changes in quantum systems. In Sect.3, we explore FSM in brief. In condensed matter physics, particularly, it is linked to BC in the quantum Hall effect and aids in detecting phase boundaries in quantum phase transitions. QGT also offers valuable insights into transport properties [39-44] providing a geometric interpretation of charge transport in systems with strong correlations or complex topological structures. In systems exhibiting superconductivity or superfluidity, QGT facilitates understanding [45] of collective excitations, such as Bogoliubov quasiparticles, which are intimately tied to the geometry of the quantum state, thereby providing insight into how macroscopic properties of superconductors or superfluid emerge from microscopic quantum states. Furthermore, QGT plays a crucial role in understanding lattice gauge theories, including spin liquids and quantum lattice models [46], examining topological order, providing insights into quantum matter, and informing on non-equilibrium systems. Recent research [45-62] has expanded topological phenomena [51,52] to non-Hermitian systems, leading to novel concepts, classifications, and applications [53-62]. This work explores the quantum geometric tensor in non-Hermitian AM models, examining the interplay of spin-orbit coupling and non-Hermiticity.

The paper is organized in the following manner: In Sect. 2, we present the model of IAM and MAM with all essential ingredients mentioned above. We obtain the corresponding band structure. In Sect.3, we explore FSM and calculate QAH conductance utilizing the model of IAM presented. We present future perspective and potential AM applications in brief in Sect. 4. The paper ends with a very brief concluding remarks in Sect. 5.

## 2. Models and method

Our investigation initially concentrates on a straightforward yet representative non-Hermitian two-dimensional model, which describes an AM on a square lattice with lattice constant a. This model serves as a foundation for examining the construction of a Hamiltonian for spinless fermions in the presence of two sublattices, A and B, on a square lattice, resulting in a fundamental 2D model of insulating AMs. For spinful fermions, it is necessary to account for the additional spin degree of freedom, yielding a four-band structure for each momentum. Conversely, spinless fermions exhibit a two-band structure. The basis for the spinful system consists of two sublattices, A and B, and two spin states, spin-up and spin-down, at each momentum point **k**. Consequently, in the momentum space, one needs to work with a four-component spinor $V = (c_{\mathbf{k},A\uparrow} \; c_{\mathbf{k},B\uparrow} \; c_{\mathbf{k},A\downarrow} \; c_{\mathbf{k},B\downarrow})^T$, where $c_{\mathbf{k},\alpha\sigma}$ are the annihilation operators for spin-up ($\sigma = \uparrow$) and spin-down ($\sigma =\downarrow$) fermions on sublattice $\alpha$ = (A,B). For the spinless case, however, $V = (c_{\mathbf{k},A} \; c_{\mathbf{k},B})^T$. We may now construct the Hamiltonian real space, considering nearest-neighbor (NN) hopping, next-nearest-neighbor (NNN) hopping, an inversion symmetry-breaking term ($M_s$) called the Semenoff-like mass [63], and periodic magnetic flux (gauge field). Furthermore, the on-site energy corresponding to the two sublattices need to be introduced. This differs for sublattices A and B and affects the energy levels associated with electrons located on each sublattice. The on-site energy difference influences the system's electronic structure, potentially leading to electron localization on one sublattice if the energy difference is substantial or causing a shift in the band structure. The introduction of a periodic magnetic flux modifies the tight-binding Hamiltonian for the system such that, while the hopping term between next-nearest-neighbor (NNN) sites remains unaffected as the closed paths of such hops enclose a complete unit cell as shown in Figure 1(a), the hopping term between NN sites acquires phase factor due to the interaction with the electromagnetic gauge field. The phase factor modifies the hopping amplitudes and affects the energy dispersion of the system. The phase can depend on both the lattice spacing and the specific configuration of the periodic magnetic flux. The on-site energy difference won't be directly affected by the flux itself. If the magnetic flux is periodic and strong enough, the system could develop Landau levels due to the magnetic field. However, for a strong periodic magnetic flux with a lattice structure, the energy levels and the corresponding wavefunctions can be more complicated and can show magnetic Bloch oscillations **[64]**, where the energy dispersion is influenced by the periodicity of the vector potential. Interestingly, the introduction of a strong periodic magnetic flux in a 2D lattice can also lead to topological effects. This magnetically induced topology is captured by the Hofstadter butterfly **[65,66]**—a fractal energy spectrum that maps out the interplay between the field and the lattice on electrons. This is especially evident in systems with a square or honeycomb lattice geometry, where the periodic vector potential causes the hopping electrons to experience a phase factor that depends on the magnetic flux, resulting in these complex energy spectra.

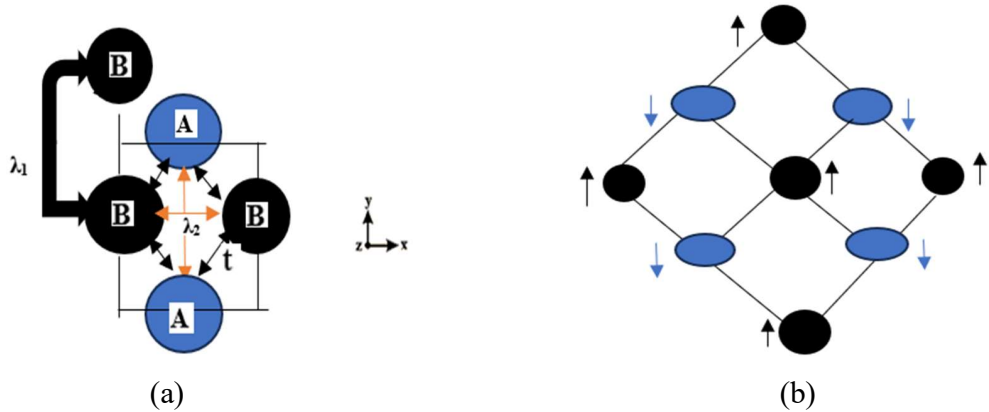

(a)            (b)

**Figure 1. (a)** (and **(b)**) The prototypical insulating (metallic) AM model to be investigated is schematically represented in Figure (a) ((b)). In Figure (a), a corresponding unit cell is clearly shown by four sites forming a square. In fact, it shows two types of sites, A and B, each attributed with an imagined *p*-orbital. The NN (*t*) and NNN ($\lambda_1$ and $\lambda_2$) hopping amplitudes are indicated in the figure **a**. The p orbitals on site A are pointing along the x-axis, and on site B, they are pointing along the y-axis. This directional dependence breaks the original rotational symmetry of the lattice from C$_4$ to C$_2$ locally at each site. Here, the B–B bonds ($\lambda_1$) are connecting upper corners of different unit cells and the A–B bonds (*t* ($\lambda_2$)) along the shorter (longer) diagonal of the same unit cell. Figure1 (b) illustrates the magnetic order on the square lattice, explicitly showing spin up (black) and down (blue) sublattices. The deformation of magnetic atom orbitals is depicted by black circles and blue ellipses. The system is also invariant under time reversal symmetry $\mathcal{R}$ and real-space mirror reflections.

In view of the observations above and the system sketched in Fig. 1(a), the Hamiltonian $\widehat{H}$ for spinless fermions – the minimalistic tight-binding model of IAM- reads

$$[\sum_{\langle m,n \rangle} t_{mn} e^{i\psi_{mn}} c_m^\dagger c_n + M_s \sum_n (-1)^{\nu_n} c_n^\dagger c_n + \sum_n (\varepsilon_n) c_n^\dagger c_n + \sum_{\langle\langle m,n \rangle\rangle} t'_{mn} c_m^\dagger c_n + \text{h.c.}] , \quad (1)$$

where $t_{mn} = t$ is the NN hopping parameter, $M_s$ is the Semenoff mass **[63]** term, $\varepsilon_n$ is the on-site energy term . The symbol $\nu_n = 0\ (1)$ depending on whether the site belongs to the A (B) sublattice. If site $n$ belongs to the sublattice A (B), the energy $\varepsilon_n = \varepsilon_A$ ( $\varepsilon_n = \varepsilon_B$). $i\gamma$ is the imaginary staggered potential added to the onsite energies leading to localized dissipation or gain at each site.

One may include the non-hermiticity by introducing the term $[i\gamma \sum_{\langle m,n \rangle}(c_m^\dagger c_n - c_n^\dagger c_m)]$, where $i\gamma$ is the imaginary potential leading to dissipation or gain. In this manner, despite being non-Hermitian, the Hamiltonian becomes PT symmetric yielding real eigenvalues under certain condition given below. We take $t'_{mn} = \lambda_1(\lambda_2)$ depending upon whether NNN hopping path joining A-A and B-B sites of different unit cells (same unit cell). In general, $\lambda_1 > \lambda_2$ as the σ bonds are usually stronger than π bonds. The orientation of the imagined *p-* orbitals may enable us to see how the σ bonds are stronger than π bonds as the distinction between σ and π bonds arises from the orientation of orbital overlap. The σ bonds are formed through the head-on overlap of atomic orbitals, such as *s-s* or *p-p*, along the axis connecting the two adjacent intercell sites. This direct overlap facilitates a stronger and more effective interaction. In contrast, π bonds result from the lateral overlap of *p*-orbitals, which are oriented perpendicular to the aforementioned axis. This sideways overlap is less effective due to this reduced orbital alignment, leading to a weaker and more indirect interaction. In the presence of a periodic magnetic flux, the NN hopping term $t_{mn}$ is modified as $t_{mn} \to t_{mn} \exp(i\psi_{mn})$, where $\psi_{mn} = \frac{e}{\hbar}\int_m^n \boldsymbol{A}.d\boldsymbol{l}$ is the path integral of the vector potential $\boldsymbol{A}$ due to the magnetic flux through the link connecting sites *m* and *n*. Here, the vector potential corresponds to an external, constant and perpendicular magnetic field B $\widehat{\boldsymbol{e_z}}$. Here $\widehat{\boldsymbol{e_z}}$ is a unit vector along the z-direction. A key inquiry is: "Does a constant perpendicular magnetic field, when expressed using the symmetric gauge or Landau gauge, generate a periodic magnetic flux through a square lattice with two sublattices?" Moreover, "Are the phase factors for NN hopping equivalent in both gauges?" The answers to these questions indicate that the magnetic flux exhibits periodicity in both gauges, significantly impacting the physical properties of the system, such as its electronic band structure. The physical outcomes, including the total flux through a unit cell, should remain the same when accurately calculated in either gauge. In essence, the results of the magnetic flux in both gauges will ultimately converge to the same physical behaviour when integrated over the entire system, as they are related by a gauge transformation. The choice of gauge (symmetric vs. Landau) alters the representation of the problem, but the physical magnetic flux through the lattice and its overall periodicity remain unchanged. Our analysis reveals a flux of ($\frac{Ba^2}{2}$), yielding the required phase factor of $\psi = \left(\frac{eBa^2}{2\hbar}\right)$. To conclude this paragraph, we note that symmetry plays a crucial role in determining the behavior of quantum matter, and its presence is ubiquitous in various lattice systems. In the present context, each lattice site is envisioned to host an imagined *p* orbital, as schematically illustrated in Figure 1(a). This envisioning aims to lower the local symmetries of the A and B sites from $C_4$ to $C_2$, preventing the lattice from reducing to a simple cubic structure with one atom in a unit cell. How does this occur? Additionally, is this procedure consistent with a model of insulating alternating magnets? If so, how? What other methods could be employed for the symmetry reduction? It may be noted in Figure 1(*a*) that the *p* orbitals on site A are pointing along the *x*-axis, and on site B, they are pointing along the *y*-axis. This directional dependence breaks the original rotational symmetry of the lattice from $C_4$ to $C_2$ locally at each site and the interactions between atoms at sites *A* and *B* are no longer symmetric in all directions. This lowering of symmetry prevents the system from reducing to a simpler cubic structure. The other methods to achieve this symmetry reduction could involve strain, anisotropic interactions, external fields, structural distortions, or interlayer couplings. These modifications can lead to an alternating magnetization pattern, characteristic of AMs, and also contribute to the insulating behavior by opening a gap in the electronic structure. Also, the symmery-lowering can lift degeneracies in the energy bands. Therefore, topologically insulating altermagnets (IAM) in square lattices exhibit a unique combination of magnetic ordering and unusual topological electronic properties, which can be described, in the basis $(c_{k,A}\ c_{k,B})^T$, by the momentum-space Haldane-like model[67,68]

$$H_{spinless}(\mathbf{k}) = h_A(\mathbf{k})\frac{(\tau_0+\tau_z)}{2} + h_B(\mathbf{k})\frac{(\tau_0-\tau_z)}{2} + d(\mathbf{k}).\boldsymbol{\tau}, \qquad (2)$$

where

$$d(\mathbf{k}) = (d_1 = -2t(\cos k_x + \cos k_y)\cos\psi + \Delta_g(\mathbf{k}), d_2 = 2t(\cos k_x + \cos k_y)\sin\psi - \gamma, d_3 = M_s),$$

$$h_{\substack{A\\B}} = \varepsilon_{\substack{A\\B}} - 2\lambda_1\big(\cos(k_x+k_y)a + \cos(k_x-k_y)a\big) - 2\lambda_2(\cos(2k_xa) + \cos(2k_ya)), \qquad (3)$$

$\tau_0$ is the 2 × 2 identity matrix, and $\boldsymbol{\tau} = (\tau_x, \tau_y, \tau_z)$ are Pauli (not spin) matrices. The Haldane theoretical framework, initially proposed for a spin-1/2 system on a honeycomb lattice, has been extended to other geometries, including square lattices under specific conditions [69]. The g-wave gap function $\Delta_g(\mathbf{k}) = \Delta_{g0}\cos(4\arctan(ak_y/ak_x))$, the specifics of which are provided below, typically couples different sublattices with opposing spin states in the case of spinful fermions. The electronic structure of the system is generally gapped at the Fermi level, and the pairing responsible for the gap is more naturally linked to opposite spins of nearest-neighbor (NN) sites due to the antiferromagnetic (AFM) nature of the system. The energy eigenvalues of $H_{spinless}(\mathbf{k})$ are given by

$$E(\mathbf{k}) = \frac{h_A(\mathbf{k}) + h_B(\mathbf{k})}{2} \pm \sqrt{J_0(k_x, k_y)} \qquad (4)$$

where $J_0(k_x, k_y) = \frac{[\varepsilon_A - \varepsilon_B]^2}{4} + d_1^2 + d_2^2 + d_3^2 + [\varepsilon_A - \varepsilon_B]d_3$. It is now easy to see that the condition for PT symmetry is $J_0(k_x, k_y) > 0$ for all values of $(k_x, k_y)$ in the first Brillouin zone. The 2D plots of the eigenvalues are presented in Figure 2(a) and (b) for the g-wave ordered insulating system as a function of the wavenumber component $ak_x$ ( The wavevector component $ak_y = 0$ and $\pi$ in Figure (a) and Figure (b), respectively.) The parameter values in the plots are $t = 1$, $\varepsilon_A = 0.41, \varepsilon_B = 0.32, M_s = 0.95, \psi = \frac{\pi}{3}, \gamma = 0.50, \mu = 0, \Delta_{g0} = 0.50, \lambda_1 = 0.01,$ and $\lambda_2 = 0.001$. The plots exhibit spectral gaps at the high symmetry points $\Gamma$ (0,0), $M$ ($\pi, \pi$), X($\pi$, 0) and Y(0, $\pi$)) of the system. The solid horizontal line represents the Fermi energy. The corresponding eigenvectors are given by $|u_\beta(k_x, k_y)\rangle = N_\beta^{-\frac{1}{2}}\Theta_\beta(\mathbf{k})$, where $N_\beta = [1 + \mathbb{R}_\beta^*\mathbb{R}_\beta]$, and $\Theta_\beta(\mathbf{k})$ is the transpose of the row vector ( 1 $\mathbb{R}_\beta$), $\beta = (1, 2)$, and $\mathbb{R}_\beta = \left(\frac{-(h_A + h_s - E(\mathbf{k}))(d_1 - id_2)}{d_1^2 + d_2^2}\right)$. Throughout the paper, we choose $t$ to be the unit of energy.

We now wish to formulate a model Hamiltonian for MnTe, an insulating AM, where the spin and the sublattice degrees of freedom will be given due consideration. Prior to this, it is essential to note that certain AFM materials exhibit a non-zero magnetic moment near absolute zero due to spin canting. This phenomenon occurs when spins are tilted by a small angle, rather than being exactly parallel. The Dzyaloshinskii-Moriya interaction (DMI) **[70-72]** is the primary interaction responsible for spin canting. The crucial feature of the DMI is that the spins are coupled in a non-collinear fashion, which is often momentum-dependent. More specifically, the DM interaction leads to an effective interaction where the spin orientation is influenced by the relative momentum between the states. The non-collinear spin coupling distinguishes the DM interaction from simpler Heisenberg-like exchange interactions and contributes to the rich behavior of systems with spin-orbit coupling. The interaction term ensures that the spin orientations on neighboring sites (or sublattices) are not aligned in a simple parallel or antiparallel manner, but instead are rotated by a relative angle that depends on the momentum difference between the states. Physically, this means that the DM interaction does not simply favor parallel or antiparallel spin alignments as would be the case in a conventional Heisenberg exchange interaction, but instead creates a chiral spin arrangement that depends on the momentum of the electrons. The spin orientations of neighboring sites are constrained by the relative orientation of their momenta, so the spin arrangement in momentum space is highly sensitive to the wavevector. It must be mentioned that the spin canting can also be influenced by anisotropic exchange interactions, crystal symmetry, and external factors such as magnetic fields or temperature.

The exchange interaction is represented by $H_{ex} = -J \sum_{ij} \mathbf{S_i} \cdot \mathbf{S_j}$ where J denotes the exchange constant, a denotes NN pairs across sublattices A and B, and $\mathbf{S_i}$ and $\mathbf{S_j}$ are spin operators at nearest neighbor sites i and j. Positive and negative J values indicate parallel and antiparallel spin alignments. The Hamiltonian of an IAM, such as MnTe **[73]**, includes NN and NNN hopping terms. Notably, MnTe's Hamiltonian features relativistic spin-orbit coupling due to its centrosymmetric nature with vanishing net magnetization. This coupling precludes spin-flip processes. In 2D systems, the relevant orbital angular momentum component is $L_z$. Spin flips often result from potential asymmetry. The material's spins alternate direction in a regular pattern, resulting in an alternating spin structure. Non-collinear alignment may appear due to DMI, leading to complex magnetic ordering **[14, 73]**. External factors, such as magnetic fields, strain, or temperature, can modify the spin configuration and contribute to spin canting. A typical Hamiltonian for the system under consideration comprises of the energy dispersion (comprising of the NN hopping parameter, the on-site energy term, and possibly the imaginary staggered potential added to the onsite energies leading to localized dissipation or gain at each site as discussed above), the exchange interaction term J, ,the DMI term, the g-wave gap function, and the relativistic spin-orbit coupling (SOC) term. In real space, the DMI contribution to an IAM Hamiltonian can be written as $H_{DMI} = \sum_{ij} D_{ij} \mathbf{S_i} \times \mathbf{S_j}$ where $D_{ij}$ is the real space DM vector that couples the spins at the NNN sites. The vector $D_{ij}$ depends on the specific material, and it is usually perpendicular to the plane of the 2D system under consideration. In momentum space, the contribution takes the form

$$\sum_{k,k'} \mathbf{D}(\mathbf{k}, \mathbf{k'}) \cdot (\mathbf{S_k} \times \mathbf{S_{-k'}}), \tag{5}$$

where $\mathbf{S_k} = \frac{1}{N^{\frac{1}{2}}} \sum_i \mathbf{S_i} e^{-i k \cdot r_i}$ and $\mathbf{S_{-k'}} = \frac{1}{N^{\frac{1}{2}}} \sum_i \mathbf{S_j} e^{i k' \cdot r_j}$ are the Fourier component of the spin operators, and $\mathbf{D}(\mathbf{k}, \mathbf{k'})$ is the Fourier transform of the DM vector $\mathbf{D_{ij}}$. Moreover, the g-wave gap function, in the context of an insulating AM like MnTe, typically couples sublattices with opposite spin states. The correct form of the g-wave order pairing corresponds to an angular momentum quantum number $l$ = 4. In momentum space, it is given by $\Delta_g(\mathbf{k}) = \Delta_{g0} \cos(4\phi_k)$, where $\phi_k$ is the azimuthal angle of the momentum vector $\mathbf{k}$. In spherical coordinates, the momentum vector $\mathbf{k}$ can be expressed as

$$\mathbf{k} = (k \sin(\theta_k) \cos(\phi_k), k \sin(\theta_k) \sin(\phi_k), k\cos(\theta_k)), \qquad (6)$$

where $\theta_k$ is the polar angle. Now, the expression $\Delta_g(\mathbf{k}) = \Delta_{g0} \cos(4\phi_k)$ ensures that this corresponds to $l$ = 4 because (a) $\phi_k$ is multiplied by 4, which means that the pairing amplitude $\Delta_g(\mathbf{k})$ has 8-fold rotational symmetry (since $4\phi_k$ has period π/2), and (b) the function $\cos(4\phi_k)$ can be expanded in terms of spherical harmonics which are eigenfunctions of the angular momentum operator with l = 4. In two dimensions, the azimuthal angle $\phi_k$ can be expressed as $\phi_k$ = arctan ($ak_y$/ $ak_x$). This enables us to write $\Delta_g(\mathbf{k}) = \Delta_{g0}$ cos (4 arctan ($ak_y$/ $ak_x$)). The Fourier transform of this expression represents the g-wave pairing order in real space, with a modulation that extends to NN sites. In contrast, as we will see below, in the metallic phase, the system has a finite density of states at the Fermi level. The g-wave gap function, in this state, may allow for different kinds of spin pairings, including those that couple same-spin states on different sublattices. Therefore, a typical Hamiltonian for the system under consideration comprises of the energy dispersion (kinetic term), the exchange interaction term, the spin-orbit coupling term, the DMI term, and the g-wave gap function. The next task is to write down the corresponding model Hamiltonian $H_{insul}$ in momentum space. As can be seen in ref. [75], in the single-electron atoms approximation $\lambda_{SOC}$ is independent of momentum.

In the basis $(c_{\mathbf{k},A\uparrow}\ c_{\mathbf{k},B\uparrow}\ c_{\mathbf{k},A\downarrow}\ c_{\mathbf{k},B\downarrow})^T$, including the DMI and the g-wave pairing, $H_{insul}(\mathbf{k})$ appears as

$$H_{insul}(\mathbf{k}) = M(\mathbf{k})\sigma_0 + f(\mathbf{k}).\boldsymbol{\sigma}, \qquad (7)$$

Where $f(\mathbf{k}) = (f_1 = \Delta_g(\mathbf{k}) + J(\mathbf{k}), f_2 = 0, f_3 = 0)$, and

$$M(\mathbf{k}) = a_1(\mathbf{k}) \frac{(\tau_0+\tau_z)}{2} + a_2(\mathbf{k}) \frac{(\tau_0-\tau_z)}{2} + t(\mathbf{k})\tau_x - t'(\mathbf{k})\tau_y + ib(\mathbf{k})\tau_z. \qquad (8)$$

The Pauli matrices $\boldsymbol{\sigma}$ and $\boldsymbol{\tau}$ are acting in the space of bands and fulfil the relation $\Theta\ \sigma/\tau\ \Theta^{-1} = -\sigma/\tau$, where $\Theta$ is the time reversal operator. The energy eigenvalues and the corresponding eigenvectors of $H_{insul}(\mathbf{k})$ together with the explicit expressions for the terms $(a_1(\mathbf{k}), a_2(\mathbf{k}), t(\mathbf{k}), t'(\mathbf{k}), b(\mathbf{k}), \Delta_g(\mathbf{k}), J(\mathbf{k}))$ are given in the Appendix. Despite being non-Hermitian, the Hamiltonian $H_{insul}(\mathbf{k})$ becomes PT symmetric yielding real eigenvalues under certain conditions. The 2D plots of the single-particle excitation spectrum $E_n$ ($n = 1,2,3,4$) given vectors of $H_{insul}(\mathbf{k})$ together with the explicit expressions for the terms $(a_1(\mathbf{k}), a_2(\mathbf{k}), t(\mathbf{k}), t'(\mathbf{k}), b(\mathbf{k}), \Delta_g(\mathbf{k}), J(\mathbf{k}))$ are given in the Appendix. Despite being non-Hermitian, the Hamiltonian $H_{insul}(\mathbf{k})$ becomes PT symmetric yielding real eigenvalues under certain conditions. The 2D plots of the single-particle excitation spectrum $E_n$ ($n = 1,2,3,4$) given by Eq. (A.2) of the $g$-wave ordered insulating system as a function of the wavenumber component $ak_x$  ( The wavevector component $ak_y = $ 0 and $\pi$ in Figure 2(c) and  Figure 2(d), respectively.) are shown in Figure 2. The numerical values of the parameters used in the plot are $t = 1$, $\varepsilon_A = 0.41$, $M_s = 0.95$, $\psi = \frac{\pi}{3}$, $\gamma = 0.23$, $J = 0.80$, $\mu = 0.50$, $\Delta_{g0} = 0.50$, $\lambda_1 = 0.01$, $\lambda_2 = 0.001$, $\lambda_{rel} = 0.01$, $D_1 = 0.62$, and $D_2 = 0.53$ in (a) and (b). The plots exhibit spectral gaps at the high symmetry points $\Gamma$ (0,0), $M$ ($\pi$, $\pi$), X($\pi$, 0) and Y(0, $\pi$)) of the system. The solid horizontal line represents the Fermi energy.

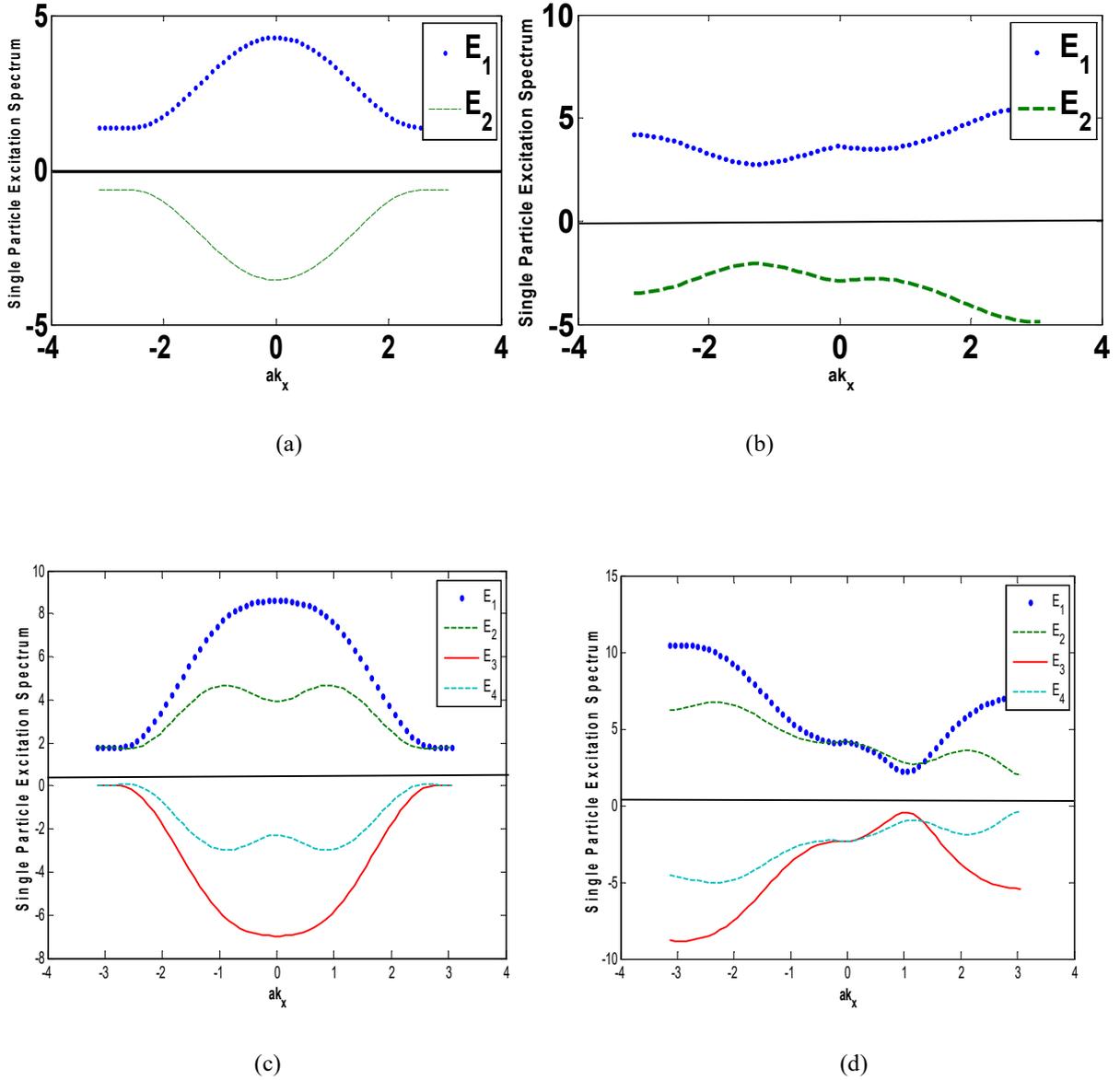

**Figure 2. (a) and (b)** The 2D plots of the eigenvalues $E(\mathbf{k}) = \frac{h_A(\mathbf{k}) + h_B(\mathbf{k})}{2} \pm \sqrt{J_0(k_x, k_y)}$ are presented in Figure 2(a) and (b) for the $g$-wave ordered insulating system as a function of the wavenumber component $ak_x$ ( The wavevector component $ak_y = 0$ and $\pi$ in Figure (a) and Figure (b), respectively. ) The parameter values in the plots are $t = 1$, $\varepsilon_A = 0.41$, $\varepsilon_B = 0.32$, $M_s = 0.95$, $\psi = \frac{\pi}{3}$, $\gamma = 0.50$, $\mu = 0$, $\Delta_{g0} = 0.50$, $\lambda_1 = 0.01$, and $\lambda_2 = 0.001$. **(c) and (d)** The 2D plots of the single-particle excitation spectrum $E_n$ ($n = 1,2,3,4$) given by Eq. (A.2) of the $g$-wave ordered insulating system as a function of the wavenumber component $ak_x$ ( The wavevector component $ak_y = 0$ and $\pi$ in Figure (a) and Figure (b), respectively. ). The numerical values of the parameters used in the plot are $t = 1$, $\varepsilon_A = 0.41$, $M_s = 0.95$, $\psi = \frac{\pi}{3}$, $\gamma = 0.23$, $J = 0.80$, $\mu = 0.50$, $\Delta_{g0} = 0.50$, $\lambda_1 = 0.01$, $\lambda_2 = 0.001$, $\lambda_{rel} = 0.01$, $D_1 = 0.62$, and $D_2 = 0.53$. The plots exhibit spectral gaps at the high symmetry points $\Gamma$ (0,0), $M$ ($\pi$, $\pi$), X($\pi$, 0) and Y(0, $\pi$)) of the system. The solid horizontal line represents the Fermi energy.

The next task is how to formulate a model 2D Hamiltonian in momentum space for an altermagnet (AM) like RuO$_2$ [11,15,16,76,77] or CrSb [78] on a square lattice, incorporating the sublattice and the spin degrees of freedom, d- or g-wave symmetry of the order parameter, the Rashba spin-orbit coupling (RSOC), and perpendicular magnetization. As a warm-up activity, we first consider a non-Hermitian, two-dimensional single orbital model characterizing MAM that describes d- /g-wave order on a square lattice with lattice constant $a$. The model provides a starting point for examining the metallic rutile RuO$_2$ [11,15,16,60,61]. The model is mathematically represented as

$$H(\mathbf{k}) = H_0(\mathbf{k}) + \Sigma(\mathbf{k} = 0), H_0(\mathbf{k}) = -[t\{\cos(ak_x) + \cos(ak_y) + \mu\}\sigma_0 + H_{RSOC}(\mathbf{k}) + [M(\mathbf{k}) + M_B]\sigma_z, \quad (9)$$

where $H_{RSOC}(\mathbf{k}) = \lambda_R[\sigma_x \sin(ak_y) - \sigma_y \sin(ak_x)]$, $\mu$ is the chemical potential, $t$ represents kinetic energy part arising from the nearest-neighbor hopping, $\lambda_{RSOC}$ characterizes Rashba spin-orbit coupling, $M(\mathbf{k}) = M_d\{\cos(ak_x) - \cos(ak_y)\}$ denotes momentum-dependent $d_{x^2-y^2}$-wave order described in terms of angular momentum $\ell=2$ harmonics in the reciprocal space; for the g-wave order we assume $M(\mathbf{k}) = 2M_g\sin(ak_x)\sin(ak_y)(\cos(ak_y) - \cos(ak_x))$. Here $\sigma_0$ and $\sigma_{x,y,z}$, respectively, are the two-by-two identity matrix and Pauli matrices. The Pauli matrices $\boldsymbol{\sigma}$ act on spins. Here, $\mathbf{M_B} = M_B \hat{e_z}$ signifies the momentum-independent magnetization component perpendicular to the plane and the unit vector $\hat{e_z}$ is in this direction. The magnetization component will arise from either the ferromagnetic insulator substrate or an external perpendicular magnetic field. AMs may involve multiple orbitals, such as d-orbitals in RuO$_2$ and g-orbitals in CrSb. Consequently, in addition to d-wave symmetry, the system may also exhibit higher-order g-wave symmetry in the order parameters. From a rigorous perspective, systems of interest that require comprehension and analysis are inherently non-closed. The Hermiticity of open systems is universally compromised, resulting in inevitable gain and loss. Notably, electron-electron, electron-impurity, and electron-photon scatterings with in electronic systems yield complex self-energies [11,15,16,60,61] for single electron states, thereby rendering their lifetimes finite. Consequently, the complex Hamiltonian can be expressed as $H(\mathbf{k}) = H_0(\mathbf{k}) + +\Sigma(\mathbf{k} = 0)$. In the case of the coupling AM with a semi-infinite FM lead (SIFML), $\Sigma = -i\Gamma\sigma_0 - i\gamma\sigma_z$ designates the self-energy term [62,63] integrating $\gamma$ for intrinsic loss. The momentum and frequency independent term $\Gamma$ in $\Sigma(= -i\Gamma\sigma_0 - i\gamma\sigma_z)$ depends on intrinsic magnetization of the lead, the hopping from the lead to the 2D AM system, and the hopping within the lead. It may be noted that $H(\mathbf{k})$ could also be written as $H(\mathbf{k}) = d_0(\mathbf{k})\sigma_0 + \mathbf{d}(\mathbf{k}) \cdot \boldsymbol{\sigma}$, where

$$d_0(\mathbf{k}) = -[t\{\cos(ak_x) + \cos(ak_y) + \mu\}] - i\Gamma \quad (10)$$

for the SIFML case. The term $\mathbf{d}(\mathbf{k}) = (d_x = \lambda_R \sin(ak_y), d_y = -\lambda_R \sin(ak_x), d_z = [M(\mathbf{k}) + M_B] - i\gamma)$. Non-Hermitian Hamiltonians differ from Hermitian ones in their potential non-diagonalizability, leading to incomplete eigenstate bases and the emergence of exceptional points (EPs)[64] at specific momenta. Open systems [64-66], inherently non-Hermitian due to environmental coupling, have driven growing interest in EPs [64-66] and non-Hermitian topological phases. We have incorporated a non-zero γ to model intrinsic loss, enabling EPs. Next, we analyze the band structure, degeneracies, and EPs. In this SIFML case, we obtain the energy eigenvalues as

$$E_1, E_2 = (A(k_x, k_y) \pm \sqrt{J_1(k_x, k_y)}),$$
$$A(k_x, k_y) = -t[\cos(ak_x) + \cos(ak_y)] - \mu,$$
$$J_1(k_x, k_y) = [M(\mathbf{k}) + M_B]^2 + \lambda_R^2 [\sin^2(ak_x) + \sin^2(ak_y)] - \gamma^2 + \Gamma^2 \quad (11)$$

considering the real part. The imaginary part yields $E_1 = A(k_x, k_y) - \frac{\gamma}{\Gamma}[M(\mathbf{k}) + M_B] = E_2$. The 2D plots of the real parts of the single-particle excitation spectra $E_1, E_2$ of the $d$-wave ( **(a)**, and **(b)** ) and the $g$-wave ( **(c)**, and **(d)** ) ordered metallic system as a function of the wavenumber component $ak_x$ ( The wavevector component $ak_y = 0$ and $\pi$ in Figures (a, c) and Figures (b, d), respectively. ) are shown in Figure 3. The numerical values of the parameters used in the plot are $t = 1, M = 0.125, \lambda_{RSOC} = 0.62, \gamma = 0.04, \mu = 0, \Delta_{d0} = 0.50 = \Delta_{g0}$, and $\Gamma = 0.7$. The solid horizontal line represents the Fermi energy. Given that the upper/lower band is partially filled in the figures showing the plots, metallicity is inherently implied. As regards the eigenvectors, the states linked to the right energy eigenvalues ($E_\alpha = E_1, E_2$) are given by $|u^{(\alpha)}(k_x, k_y)\rangle = \mathcal{M}^{(\alpha)-\frac{1}{2}}\phi_\alpha(\mathbf{k})$, where $\phi_\alpha(\mathbf{k}) = (\psi_1^{(\alpha)} \quad \psi_2^{(\alpha)})^T$. The matrix components

$$\psi_1^{(1)} = -\lambda_R[\sin(ak_y) + i\sin(ak_x)], \psi_2^{(1)} = (M(\mathbf{k}) + M_B - (E_1 + t(\mathbf{k})) - i\gamma_1),$$
$$\psi_1^{(2)} = (M(\mathbf{k}) + M_B - (E_2 + t(\mathbf{k})) - i\gamma_2), \psi_2^{(2)} = -\lambda_R[\sin(ak_y) - i\sin(ak_x)], (12)$$

$\gamma_1 = \gamma + \Gamma, \gamma_2 = \gamma - \Gamma$, and $\mathcal{M}^{(\alpha)}$ is to defined by a bi-orthonormality condition given below.

At EPs multiple eigenvalues and their corresponding eigenvectors coalesce. This coalescence leads to linear dependence among the eigenvectors, rendering the Hamiltonian non-diagonalizable and imparting a distinct algebraic signature indicative of an EP. Unlike Hermitian systems, non-Hermitian frameworks lack conventional orthogonality among eigenvectors. This limitation is addressed through the use of bi-orthogonal bases, which help restore analytical tractability. However, this approach introduces ambiguity in distinguishing between quantum states and observables, as observables no longer retain a straightforward Hermitian interpretation. To recover a probabilistic framework, a modified inner product is employed, defined via a non-trivial metric operator η. This operator typically contains off-diagonal elements, signifying long-range correlations between spatially separated components. Such behaviour reflects the inherently delocalized nature of the bi-orthogonal basis in non-Hermitian systems. In this context, left and right eigenvectors are introduced to generalize the concept of orthogonality. Specifically, the right eigenvectors satisfy $H(\mathbf{k})|u^{(j)}\rangle = E_j |u^{(j)}\rangle$ (right eigenvectors), and $H(\mathbf{k})^\dagger |v^{(j)}\rangle = E_j^* |v^{(j)}\rangle$ or, $\langle v^{(j)}|H(\mathbf{k}) = E_j\langle v^{(j)}|$ (left eigenvectors). Here, the eigenvalues $E_j$ and the corresponding eigenvectors ($|u^{(j)}\rangle, \langle v^{(j)}|$) of the Hamiltonian $H(\mathbf{k})$ are indexed by $j = 1, 2$. These eigenvectors do not exhibit standard orthonormality but instead form a bi-orthonormal system satisfying $\langle v^{(i)}|u^{(j)}\rangle = \delta_{ij}$ [63,64]. Notably, the Gram matrices constructed from the right and left eigenvectors are not generally inverses of one another, i.e., $(\langle u^{(i)}|u^{(j)}\rangle) \neq (\langle v^{(i)}|v^{(j)}\rangle)^{-1}$.

However, inverse relationships can be established if the eigenvectors are explicitly bi-orthonormalized. A key quantitative measure of EPs is the Phase rigidity, defined as

$$P_j = |\langle v^{(j)}|u^{(j)}\rangle|/|\langle u^{(j)}|u^{(j)}\rangle|. \tag{13}$$

As the system approaches an EP, $P_j \to 0$ for the coalescing states, making phase rigidity a reliable indicator of EP behaviour. The metallic AMs naturally support EPs due to their band structure and spin-split conduction bands, which allow for non-Hermitian perturbations (like gain/loss or coupling to reservoirs). Figures 3e and 3f present the results of our exploration into this issue based on (12). The function $P_1(k_x, k_y)$ [$P_2(k_x, k_y)$] is plotted in Fig. 3e (3(f)) for the $d$-wave ordered metallic system as a function of the wavenumber component $ak_x$. The wavevector component $ak_y = 0$ and $\pi$ in Fig. 3e and Fig. 3f, respectively. The numerical values of the parameters used in the plot are $t = 1$, $M = 0.125$, $\lambda_{RSOC} = 0.62$, $\gamma = 0.032$, $\mu = 0$, $\Delta_{d0} = 0.14 = \Delta_{g0}$, and $\Gamma = 0.8$. The solid horizontal line corresponds to $P_j(k_x, k_y) = 0$. The blue colored arrows in these figures indicate the location of the EPs. As regards the physical manifestations of exceptional points, these points in non-Hermitian systems, including insulating AMs, often lead to unique physical phenomena, such as non-Hermitian skin effects, unusual transport properties, and anomalous behaviour in response to external perturbations.

To construct a somewhat realistic Hamiltonian in real and momentum spaces with hopping terms between nearest and next-nearest neighbors, it is essential to address the key features of this system: the square lattice structure with two sublattices, the orbital nature of the electrons, the presence of magnetization, and the wave-ordering symmetries. For a 2D AM, the system is often described by two sublattices (A and B), allowing for alternating spin configurations (see Figure 1(b)). These sublattice sites facilitate the capture of alternating magnetic order characteristic of AM. The magnetization, assumed to be perpendicular to the 2D plane, influences the spin degrees of freedom. This term in a model Hamiltonian accounts for spin polarization and is typically described by an exchange interaction term coupling the spin degrees of freedom on different sublattices (A and B). The RSOC accounts for spin-orbit coupling effects, involving momentum and spin interactions. In the specific case of a square lattice with two sublattices (A and B), the Rashba term couples the spin states on sublattice A with those on sublattice B, mixing the spin-up and spin-down states between the two sublattices. The spin-orbit coupling arises from the asymmetry of the system, where inversion symmetry is broken, such as in the presence of an electric field, surface states, or spin-polarized materials. In the context of d-wave pairing, the order parameter typically facilitates coupling between electrons residing on distinct sublattices, namely A and B, which exhibit opposite spin states. This phenomenon arises from the propensity of the d-wave gap function to undergo a sign change across the sublattices, thereby promoting pairing between electrons situated on these disparate sublattices. The favored pairing mechanism involves opposite-spin pairing between the sublattices, which is further reinforced by the spin-polarized nature of the system. In contrast, the g-wave order [78] typically gives rise to coupling between electrons within the same sublattice, either A-A or B-B, rather than between distinct sublattices. This behavior can be attributed to the fact that the gap function does not exhibit the same sign change as the d-wave order, and the anticipated spin texture is expected to yield same-spin state pairing within each sublattice, resulting in either both spin-up or both spin-down states. These symmetries are consistent with the behavior of antiferromagnetic materials when subjected to various pairing mechanisms, where the magnetization influences the spin structure in a manner that promotes either opposite-spin or same-spin pairings, depending on the nature of the order parameter. By integrating these elements, the comprehensive Hamiltonian in real space for a metallic AM on a square lattice with two sublattices A and B, incorporating spin degrees of freedom, nearest-neighbor hopping, d-wave or g-wave order, spin-orbit coupling, and magnetization, can be formulated. Based these observations and the system shown in Figure 1(b), the Hamiltonian $H$ for spinful fermions – the tight-binding model of MAM – is represented as

$$[-t \sum_{\langle i,j \rangle, \sigma\sigma'} c_{i\sigma}^\dagger c_{j\sigma'} + i\gamma \sum_{i,\sigma} \theta_i c_{i\sigma}^\dagger c_{i\sigma} + \sum_{i,\sigma} \boldsymbol{M}_i \cdot \widehat{\boldsymbol{e}_z} c_{i\sigma}^\dagger c_{i\sigma} + \lambda_R \sum_{\langle i,j \rangle, \sigma\sigma', \sigma \neq \sigma'} c_{i,\sigma}^\dagger (a\boldsymbol{d}_{ij} \times \boldsymbol{\sigma}) \cdot \widehat{\boldsymbol{e}_z} \, c_{j,\sigma'} + H.c.] + F_{d/g}, \quad (14)$$

where $\langle i,j \rangle$ denotes NN pair across sublattices A and B, $i\gamma$ is the imaginary staggered potentials (ISP) representing the non-Hermitian dissipation/gain term, $\theta_i = +1(-1)$ for the sublattice A(B), $\widehat{\boldsymbol{e}_z}$ is the unit vector along z-axis, and the term $F_{d/g}$ corresponds to the d-wave (or g-wave) pairing in the real space. While $t$ is the NN hopping and $\boldsymbol{\sigma}$ the Pauli spin matrices, $\boldsymbol{M}_i$ is the local magnetization at the site $i$ and $\widehat{\boldsymbol{e}_z}$ is a unit vector perpendicular to the two-dimensional plane. If we write for simplicity $\boldsymbol{M}_{up} = M\widehat{\boldsymbol{e}_z}$, and $\boldsymbol{M}_{down} = -M\widehat{\boldsymbol{e}_z}$, the third term above reduces to $[M \sum_{i,\sigma} \alpha_i c_{i\sigma}^\dagger c_{j\sigma}]$, where $\alpha_i = +1(-1)$ for the spin up(down) states. As regards the fourth term, $\lambda_R$ is the spin-orbit coupling strength, $\boldsymbol{d}_{ij}$ is the unit vector that points from site $i$ to $j$. For rightward (upward) hopping (see Figure 1(a)), $\boldsymbol{d}_{ij} = \widehat{\boldsymbol{e}_x}$ $(\boldsymbol{d}_{ij} = \widehat{\boldsymbol{e}_y})$. We now consider the term $F_{d/g}$, where

$$F_d = [\sum_{\langle i,j \rangle, \sigma\sigma', \sigma \neq \sigma'} \Delta_{\langle i,j \rangle}^{(d)} c_{i\sigma}^\dagger c_{j\sigma'} + H.c.],$$

$$F_g = \sum_{\langle\langle i,j \rangle\rangle, \sigma\sigma', \sigma = \sigma'} \Delta_{\langle\langle i,j \rangle\rangle}^{(g)} c_{i\sigma}^\dagger c_{j\sigma'}. \quad (15)$$

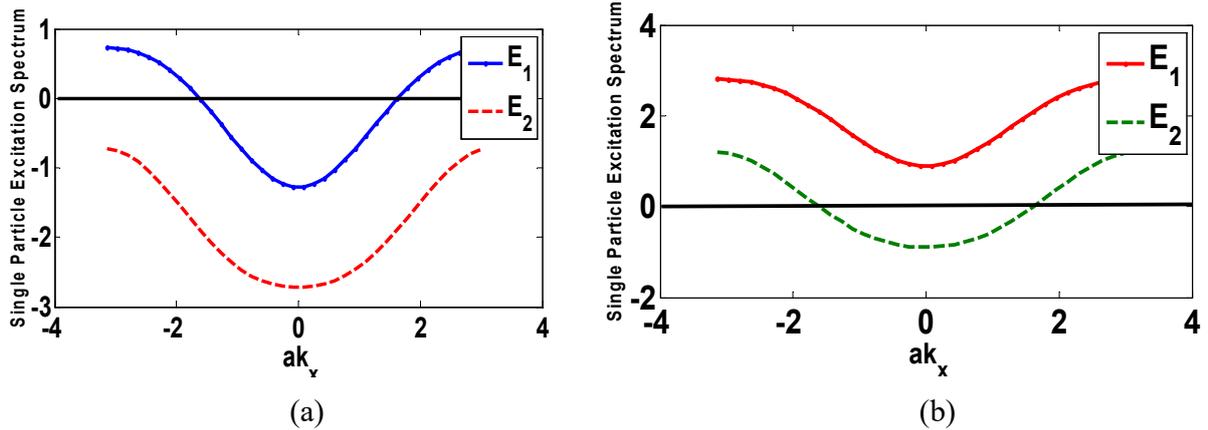

(a)          (b)

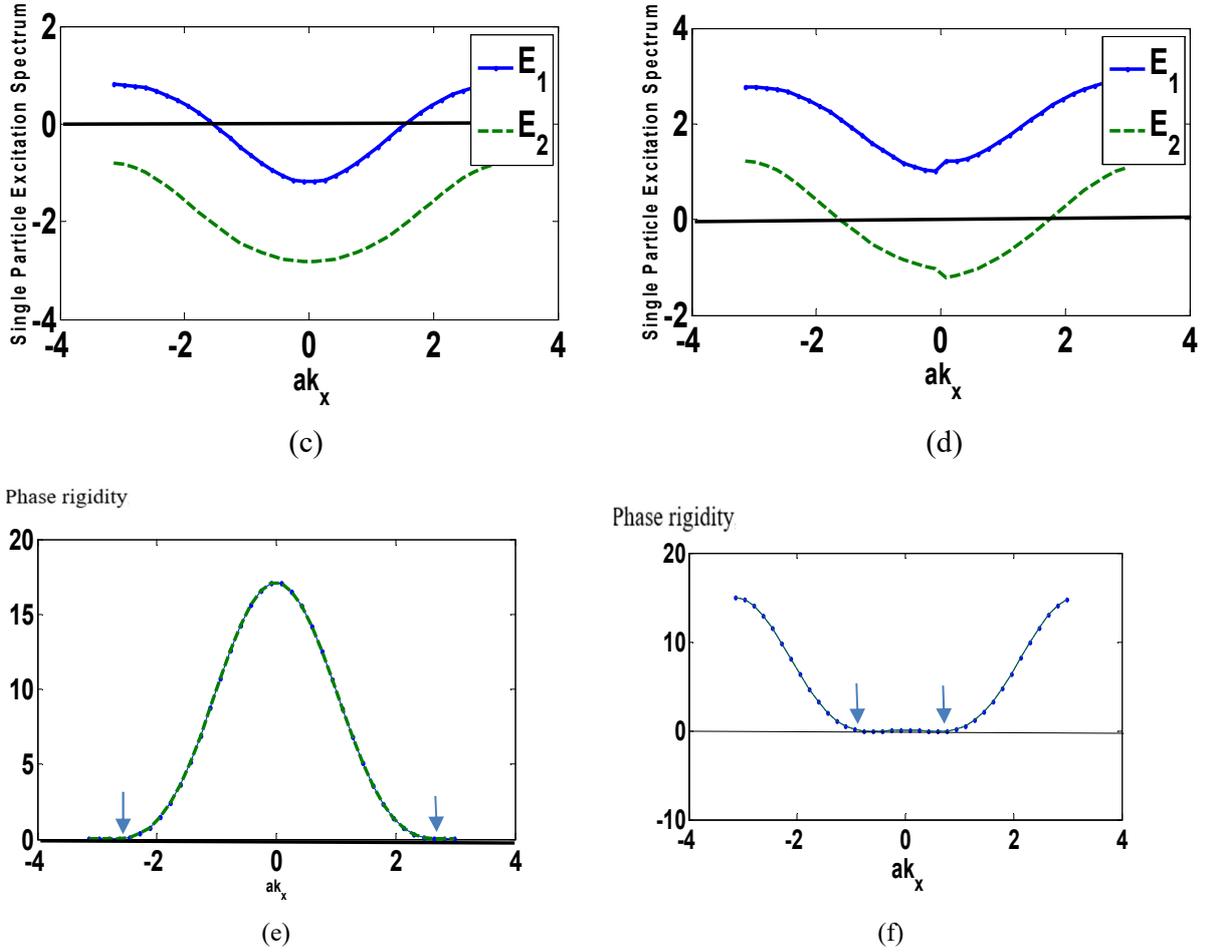

**Figure 3.** **(a), (b), (c) and (d)** The 2D plots of the single-particle excitation spectra $E_1, E_2 = (A(k_x, k_y) \pm \sqrt{J_1(k_x, k_y)})$ of the *d*-wave ( **(a),** and **(b)** ) and the *g*-wave ( **(c),** and **(d)** ) ordered systems as a function of the wavenumber component $ak_x$ ( The wavevector component $ak_y = 0$ and $\pi$ in Figures (a, c) and Figures (b, d), respectively. ). The numerical values of the parameters used in the plot are $t = 1$, $M = 0.125$, $\lambda_{RSOC} = 0.62, \gamma = 0.04, \mu = 0$, $\Delta_{d0} = 0.50 = \Delta_{g0}$, and $\Gamma = 0.7$. The solid horizontal line represents the Fermi energy. **(e) (f)** The 2D plots of the phase rigidity is represented in these figures for the *d*-wave ordered metallic system as a function of the wavenumber component $ak_x$. The wavevector component $ak_y = 0$ and $\pi$ in Fig. e and Fig. f, respectively. The numerical values of the parameters used in the plot are $t = 1$, $M = 0.125$, $\lambda_{RSOC} = 0.62, \gamma = 0.032, \mu = 0, \Delta_{d0} = 0.14 = \Delta_{g0}$, and $\Gamma = 0.8$. The solid horizontal line corresponds to the phase rigidity = 0.

The modulation factor $\Delta_{\langle i,j \rangle \atop \langle\langle i,j \rangle\rangle}^{\binom{d}{g}} = \left(\frac{1}{N}\right) \sum_k \Delta_{d \atop g}(\boldsymbol{k}) \, exp \, (i\boldsymbol{k}.\boldsymbol{r})$ where N is the number of sites. To obtain $\Delta_{\langle i,j \rangle}^{(d)}$, we need to perform a Fourier transform of $\Delta_d(\boldsymbol{k}) = \Delta_{d0}\{\cos(ak_x) - \cos(ak_y)\}$ which denotes momentum-dependent $d_{x^2-y^2}$-wave order described in terms of angular momentum $\ell=2$ harmonics in the reciprocal space. The Fourier transform of $\cos(ak_x)$ and $\cos(ak_y)$ yields delta functions: $\cos(ak_x) \to \delta(ak_x \pm 1)$ and $\cos(ak_y) \to \delta(ak_y \pm 1)$. Therefore, the modulation factor in real space for the d-wave pairing is $\Delta_{d0}[\delta(ak_x \pm 1) - \delta(ak_y \pm 1)]$. This expression represents a modulation of the pairing amplitude that alternates between neighbouring sites along the x and y directions. In momentum space, the g-wave gap function is given by $\Delta_g(\boldsymbol{k}) = \Delta_{g0} \cos(4 \arctan(ak_y/ak_x))$, where the function allows for the coupling between same-spin states on A-A and B-B sublattices. The Fourier transform of $\Delta_g(\boldsymbol{k})$ represents the g-wave pairing order in real space, with a modulation that extends to NNN sites. In view of the discussion above, upon taking Fourier transform, while the momentum space Hermitian Hamiltonian matrix for the d-wave pairing may be written as $H_d(\mathbf{k})$ in the basis $(c_{\mathbf{k},A\uparrow}\ c_{\mathbf{k},B\downarrow}\ c_{\mathbf{k},A\downarrow}\ c_{\mathbf{k},B\downarrow}\ c_{\mathbf{k},B\uparrow})^T$, for the g-wave pairing the matrix appears as $H_g(\mathbf{k})$ in the same basis. These matrices, including the corresponding energy eigenvalues ($E_{d,n}$ and $E_{g,n}$) and the eigenvectors are given in Appendix A. The 2D plots of the single-particle excitation spectrum $E_{d,n}$ ($n = 1,2,3,4$) given by Eq. (A.5), obtained from the *d*-wave ordered model, as a function of the wavenumber component $ak_x$ (The wavevector component $ak_y = 0$ and $\pi$ in Figure (a) and Figure (b), respectively.) are shown in Figure 5. The numerical values of the parameters used in the plot are $t = 1, \gamma = 0, M = 0.125, \mu = 0, \Delta_{d0} = 0.14, \lambda_R = 0.62, \lambda_1 = 0.02$, and $\lambda_2 = 0.002$. While the plot (c) exhibits the Kramers degeneracies at the high symmetry points (HSPs) X($\pm\pi$, 0), the plot (d) displays the degeneracies at the HSPs Y(0, $\pm\pi$), and M($\pm\pi, \pm\pi$). In Fig.4, the 2D plots of the single-particle excitation spectrum $E_{g,n}$ ($n = 1,2,3,4$) given by Eq. (36) in the Appendix A, obtained from the g-wave ordered model, as a function of the wavenumber component $ak_x$ (The wavevector component $ak_y = 0$ and $\frac{\pi}{4}$ in Figure (c) and Figure (d), respectively) are shown. The plot (d) highlights degeneracies at high-symmetry points $\Gamma(0,0)$ and X($\pm\pi$,0), using parameters: $t = 1, \gamma = 0, M = 0.125, \mu = 0, \Delta_{g0} = 0.14, \lambda_R = 0.63, \lambda_1 = 0.02,$ and $\lambda_2 = 0.002$. In AMs, non-Hermitian Hamiltonians with symmetries like time-reversal or particle-hole influence topological features such as the quantum geometric tensor. In the next section, Berry curvature is derived from eigenvectors (see Appendix A), followed by Chern number computation. Results hint at a possible quantum anomalous Hall (QAH) phase.

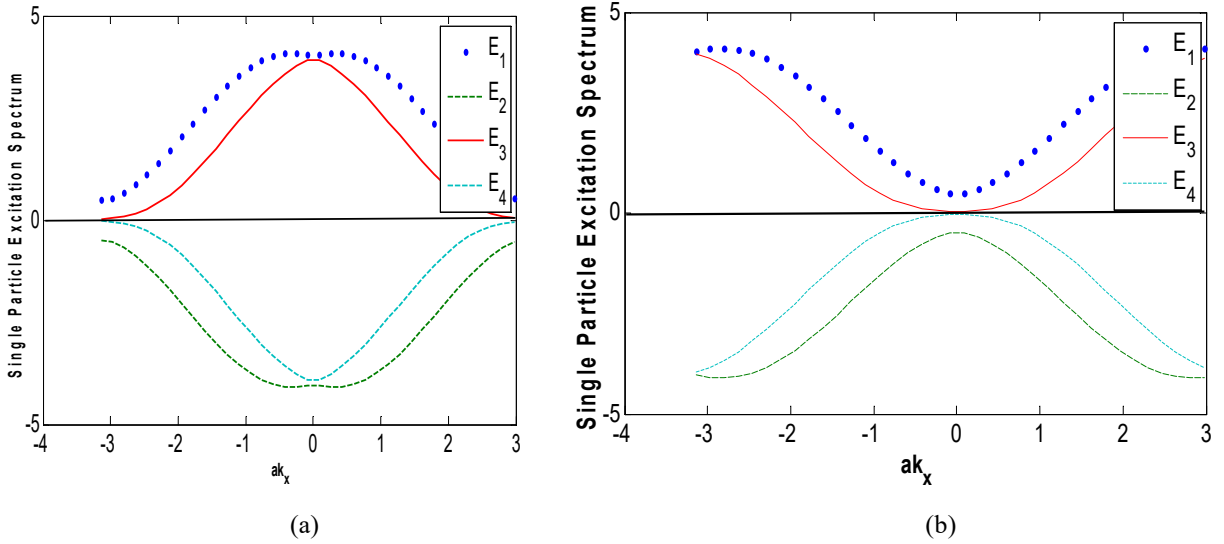

(a)  (b)

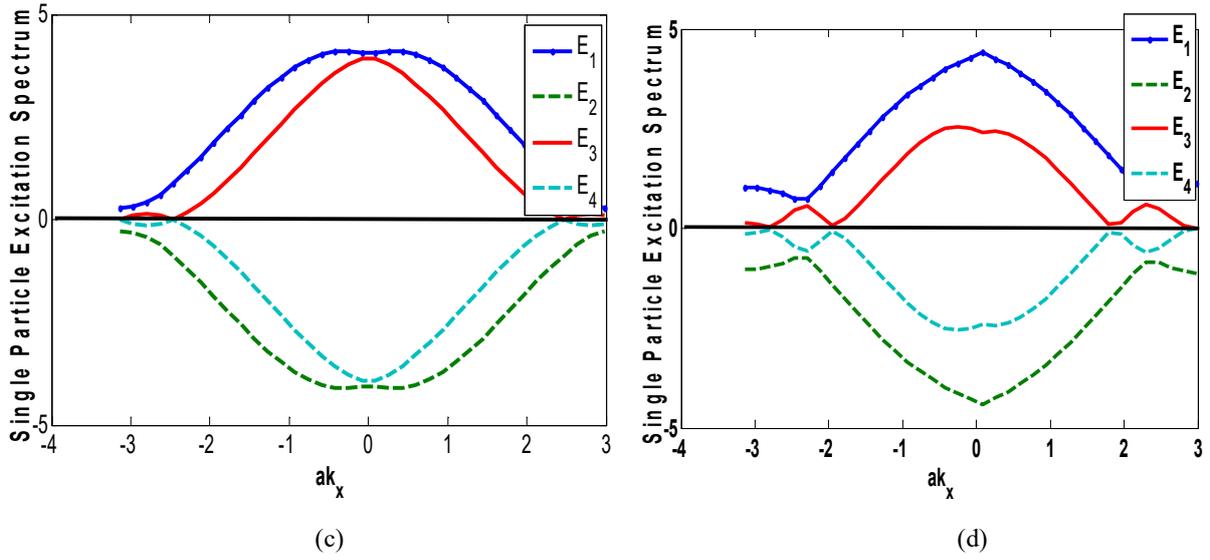

|  (c)  |  (d)  |

**Figure 4. (a) and (b)** The 2D plots of the single-particle excitation spectrum $E_{d,n}$ ($n = 1,2,3,4$) given by Eq. (A.5), obtained from the $d$-wave ordered model of AM, as a function of the wavenumber component $ak_x$ (The wavevector component $ak_y = 0$ and $\pi$ in Figure (a) and Figure (b), respectively.). The numerical values of the parameters used in the plot are $t = 1, \gamma = 0, M = 0.125, \mu = 0, \Delta_{d0} = 0.14, \lambda_R = 0.62, \lambda_1 = 0.02,$ and $\lambda_2 = 0.002$. While the plot (a) exhibits the Kramers degeneracies at the high symmetry points (HSPs) X($\pm\pi$, 0), the plot (b) displays the degeneracies at the HSPs Y(0, $\pm\pi$), and M ($\pm\pi, \pm\pi$). **(c) and (d)** The 2D plots of the single-particle excitation spectrum $E_{g,n}$ ($n = 1,2,3,4$) given by Eq. (A.6), obtained from the $g$-wave ordered model of g AM, as a function of the wavenumber component $ak_x$ (The wavevector component $ak_y = 0$ and $\pi/4$ in Figure (c) and Figure (d), respectively.). The numerical values of the parameters used in the plot are $t = 1, \gamma = 0, M = 0.125, \mu = 0, \Delta_{g0} = 0.14, \lambda_R = 0.63, \lambda_1 = 0.02,$ and $\lambda_2 = 0.002$. The plot (c) shows the degeneracies at the HSPs $\Gamma$ (0, 0) and X($\pm\pi$, 0). The solid horizontal line represents the Fermi energy.

## 3. Quantum Geometric Tensor

As already mentioned, quantum geometric tensor (QGT)[79-81] $G_{\mu\nu}$ is a matrix that captures the geometry of the quantum wavefunctions in parameter space (here, momentum space $k$) and is defined as

$$G_{n,\mu\nu}(\boldsymbol{k},\lambda) = \left\langle \frac{\partial u_{n,k}(\lambda)}{\partial k_\mu} \bigg| \frac{\partial u_{n,k}(\lambda)}{\partial k_\nu} \right\rangle - \left\langle \frac{\partial u_{n,k}(\lambda)}{\partial k_\mu} \bigg| u_{n,k}(\lambda) \right\rangle \left\langle u_{n,k}(\lambda) \bigg| \frac{\partial u_{n,k}(\lambda)}{\partial k_\nu} \right\rangle \quad (16)$$

for a Hermitian system where the symbol $|u_{n,k}(\lambda)\rangle$, which smoothly depends on the $N$-dimensional parameter $\lambda = (\lambda_1, ..., \lambda_N)$, stands for the nth eigenenergy of a quantum Hamiltonian $H$ and for a given momentum $\boldsymbol{k}$. It is important to note that, in momentum space, the derivatives of the eigenstates or wavefunctions would be taken with respect to momentum components, not the parameters $\lambda_1, ..., \lambda_N$, unless those parameters also influence the momentum in some way. The inner product is taken in the space of the eigenfunctions. The real part of the QGT is the quantum metric tensor (QMT) $g_{n,\mu\nu}(\mathbf{k},\lambda)$ = Re $G_{n,\mu\nu}(\mathbf{k},\lambda)$ which defines the distance between two quantum states, $ds^2 = g_{n,\mu\nu}(\lambda)\, d\lambda_\mu d\lambda_\nu$. Its imaginary part corresponds to the Berry curvature $\Omega_{n,\mu\nu}(\mathbf{k},\lambda) = -2\, \text{Im}\, G_{n,\mu\nu}(\mathbf{k},\lambda)$. BC is related to the topological properties of the system and can give insight into phenomena such as the anomalous Hall effect in systems with broken TRS. The QMT $g_{\mu\nu}(\mathbf{k}) = \frac{1}{2}\left(G_{\mu\nu}(\mathbf{k}) + G^*_{\nu\mu}(\mathbf{k})\right)$ is the real, symmetric part of QGT. This part of the tensor describes the local curvature of the quantum state in parameter space and very crucial for understanding non-equilibrium properties of the system, such as the response to external perturbations. The BC and QMT can also be written in terms of the derivatives of the Hamiltonian

$$\Omega_{\alpha,\mu\nu}(\mathbf{k}) = i\left[\begin{array}{c}\sum_{\beta\neq\alpha}\{(E_\alpha(\mathbf{k}) - E_\beta(\mathbf{k}))^{-2}\left\langle u_\alpha(\mathbf{k})\left|\frac{\partial H}{\partial k_\mu}\right|u_\beta(\mathbf{k})\right\rangle\left\langle u_\alpha(\mathbf{k})\left|\frac{\partial H}{\partial k_\nu}\right|u_\beta(\mathbf{k})\right\rangle - \\ (E_\alpha(\mathbf{k}) - E_\beta(\mathbf{k}))^{-2}\left\langle u_\alpha(\mathbf{k})\left|\frac{\partial H}{\partial k_\nu}\right|u_\beta(\mathbf{k})\right\rangle\left\langle u_\alpha(\mathbf{k})\left|\frac{\partial H}{\partial k_\mu}\right|u_\beta(\mathbf{k})\right\rangle\}\end{array}\right], \quad (17)$$

$$g_{\alpha,\mu\nu}(\mathbf{k}) = Re\left[\sum_{\beta\neq\alpha}\{(E_\alpha(\mathbf{k}) - E_\beta(\mathbf{k}))^{-2}\left\langle u_\alpha(\mathbf{k})\left|\frac{\partial H}{\partial k_\mu}\right|u_\beta(\mathbf{k})\right\rangle\left\langle u_\alpha(\mathbf{k})\left|\frac{\partial H}{\partial k_\nu}\right|u_\beta(\mathbf{k})\right\rangle\right]. \quad (18)$$

Additionally, the QMT is linked to the fidelity $F$, a metric that evaluates the overlap between two quantum states, $F(|u\rangle, |v\rangle) = |\langle u|v\rangle|^2$, and has been instrumental in revealing spectral degeneracies and analyzing phase transitions. The Quantum Geometric Tensor (QGT) is related to the fidelity between the quantum states $|u_{n,k}(\lambda)\rangle$ and $|u_{n,k}(\lambda + d\lambda)\rangle$ expanded to second order in $d\lambda$. Specifically, the fidelity $F(|u_{n,k}(\lambda)\rangle, |u_{n,k}(\lambda + d\lambda)\rangle) = \langle u_{n,k}(\lambda)|u_{n,k}(\lambda + d\lambda)\rangle\langle u_{n,k}(\lambda + d\lambda)|u_{n,k}(\lambda)\rangle$, which can be approximated as $(1 - \varsigma_F d\lambda_\mu d\lambda_\nu + ...)$, where $\varsigma_F$ is the fidelity susceptibility and an alternative expression of the QMT.

In a non-Hermitian system, QGT may still be defined but one has to account for the fact that the eigenstates of the non-Hermitian Hamiltonian are generally not orthonormal in the usual sense. To make the spectral and topological analysis consistent, one typically normalizes these eigenvectors to satisfy $\langle u^L(\lambda)|u^R(\lambda)\rangle = \delta_{R,L}$ as mentioned above. For such systems, the BC and QMT must be modified by considering the left and the right eigenstates. In fact, the interplay between left and right eigenstates yields four distinct combinations - left-right (LR), right-left (RL), left-left (LL), and right-right (RR) - each providing a unique definition of BC

$$\Omega^{\alpha\beta}_{n,\mu\nu}(\mathbf{k},\lambda) = i\left(\left\langle\frac{\partial u^\alpha_{n,k}(\mathbf{k},\lambda)}{\partial k_\mu}\left|\frac{\partial u^\beta_{n,k}(\mathbf{k},\lambda)}{\partial k_\nu}\right\rangle\right) - \left(\left\langle\frac{\partial u^\alpha_{n,k}(\mathbf{k},\lambda)}{\partial k_\nu}\left|\frac{\partial u^\beta_{n,k}(\mathbf{k},\lambda)}{\partial k_\mu}\right\rangle\right)\right) \quad (19)$$

where $\alpha, \beta = L, R$ represent the left and right eigenstates, subject to the normalization condition $\langle u^L(\lambda)|u^R(\lambda)\rangle = \delta_{R,L}$. Notably, these four Berry curvatures exhibit local differences, but ultimately yield the same Chern number upon integration. In a similar vein, the extension of generalized QMT, and fidelity to non-Hermitian systems can be achieved through diverse approaches.

As discussed above, the definition of QGT corresponds to the distance between quantum states [18]. In a normed vector space in quantum mechanics, the distance between two vectors is defined using a norm determined by the inner product of the quantum states. Consequently, the two distinct definitions arise from different methods of defining the inner product in non-Hermitian quantum mechanics. In the first approach, the inner product is assumed to be identical to that in Hermitian systems. In this particular scenario, upon restoring the gauge invariance, the right-right (RR) QGT

$$G_{n,\mu\nu}^{RR}(\mathbf{k},\lambda) = \left\langle \frac{\partial u_{n,k}^R(\mathbf{k},\lambda)}{\partial k_\mu} \middle| \frac{\partial u_{n,k}^R(\mathbf{k},\lambda)}{\partial k_\nu} \right\rangle - \left\langle \frac{\partial u_{n,k}^R(\mathbf{k},\lambda)}{\partial k_\mu} \middle| u_{n,k}^R(\mathbf{k},\lambda) \right\rangle \left\langle u_{n,k}^R(\mathbf{k},\lambda) \middle| \frac{\partial u_{n,k}^R(\mathbf{k},\lambda)}{\partial k_\nu} \right\rangle \quad (20)$$

emerges, characterized as a Hermitian tensor. The imaginary component of this tensor is equivalent to the RR Berry curvature $\Omega_{n,\mu\nu}^{RR}(\mathbf{k},\lambda) = -2\text{Im}\, G_{n,\mu\nu}^{RR}(\mathbf{k},\lambda)$. Therefore, it logically follows that the RR QMT is defined by its real part $g_{n,\mu\nu}^{RR}(\mathbf{k},\lambda) = \text{Re}\, G_{n,\mu\nu}^{RR}(\mathbf{k},\lambda)$ **[79-81]**.

The left-right (LR) QGT, given by

$$G_{n,\mu\nu}^{LR}(\mathbf{k},\lambda) = \left\langle \frac{\partial u_{n,k}^L(\mathbf{k},\lambda)}{\partial k_\mu} \middle| \frac{\partial u_{n,k}^R(\mathbf{k},\lambda)}{\partial k_\nu} \right\rangle - \left\langle \frac{\partial u_{n,k}^L(\mathbf{k},\lambda)}{\partial k_\mu} \middle| u_{n,k}^R(\mathbf{k},\lambda) \right\rangle \left\langle u_{n,k}^L(\mathbf{k},\lambda) \middle| \frac{\partial u_{n,k}^R(\mathbf{k},\lambda)}{\partial k_\nu} \right\rangle, \quad (21)$$

is characterized by non-Hermiticity, as observed in references **[79-81]**. Its anti-symmetric component corresponds to the LR Berry curvature $\Omega_{n,\mu\nu}^{LR}(\mathbf{k},\lambda) = i(G_{n,\mu\nu}^{LR}(\mathbf{k},\lambda) - G_{n,\nu\mu}^{LR}(\mathbf{k},\lambda))$, exhibiting complex values, as investigated in **[79-81]**. Within the realm of $PT$-symmetric or pseudo-Hermitian systems, the LR QMT is identified as the real and symmetric part of the LR QGT $g_{n,\mu\nu}^{LR}(\mathbf{k},\lambda) = Re(\frac{1}{2})(G_{n,\mu\nu}^{LR}(\mathbf{k},\lambda) + G_{n,\mu\nu}^{RL}(\mathbf{k},\lambda))$. However, the precise connection between LR QGT and LR QMT in general non-Hermitian systems remains to be elucidated. Ref. [29] presents three alternative generalizations: $g_{n,\mu\nu}^{LR}(\mathbf{k},\lambda) = (\frac{1}{2})(G_{n,\mu\nu}^{LR}(\mathbf{k},\lambda) + G_{n,\nu\mu}^{LR}(\mathbf{k},\lambda))$ (symmetric part), $g_{n,\mu\nu}^{LR}(\mathbf{k},\lambda) = Re(G_{n,\mu\nu}^{LR}(\mathbf{k},\lambda))$ (real part), and $g_{n,\mu\nu}^{LR}(\mathbf{k},\lambda) = Re(\frac{1}{2})(G_{n,\mu\nu}^{LR}(\mathbf{k},\lambda) + G_{n,\mu\nu}^{RL}(\mathbf{k},\lambda))$ (real and symmetric part). In a similar vein, diverse approaches have been employed to extend fidelity and fidelity susceptibility to non-Hermitian systems, as reported in **[82]**. Investigating this issue for the system in question becomes a crucial objective for future research.

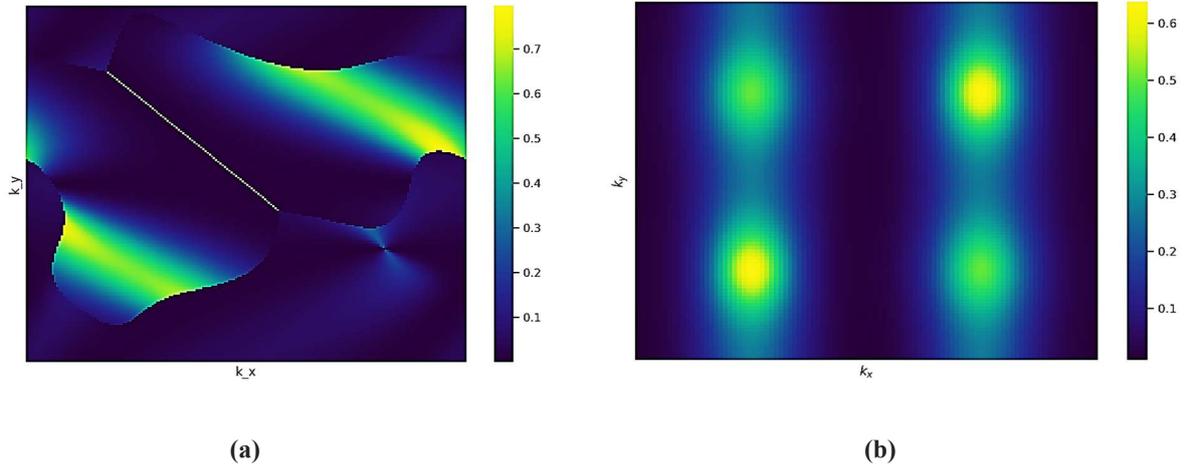

(a)          (b)

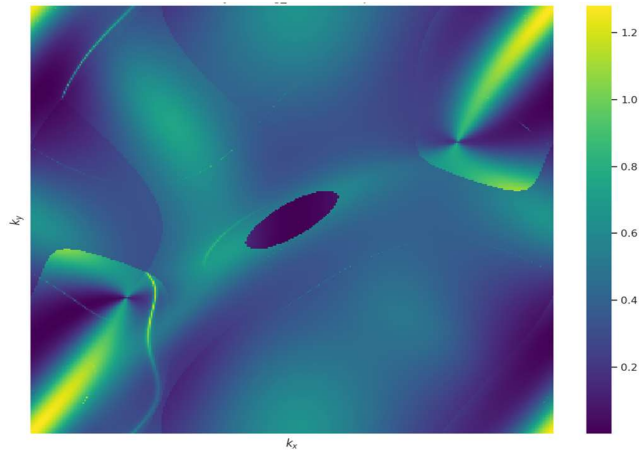

(c)

**Figure 5.** (a) The contour plot of $g^{RR}_{n,\mu\nu}(k_x, k_y) = ReG^{RR}_{n,\mu\nu}(k_x, k_y)$ for the lowest band over $[-\pi,\pi]\times[-\pi,\pi]$, assuming half-filling, using the parameters values $t = 1$, $\varepsilon_A = 0.41$, $M_s = 0.95$, $\psi = \frac{\pi}{3}, \gamma = 0.23, J = 0.80, \Delta_{g0} = 0.50$, $\lambda_1 = 0.01, \lambda_2 = 0.001$, $\lambda_{rel} = 0.01$, $D_1 = 0.62$, and $D_2 = 0.53$. **(b)** The refined contour plot of $g^{RR}_{n,\mu\nu}(k_x, k_y)$ summed over the occupied bands across $[-\pi,\pi]\times[-\pi,\pi]$, assuming half-filling, using the same parameters values. **(c)** The contour plot of $g^{RR}_{n,\mu\nu}(k_x, k_y)$ summed over the occupied bands across $[-\pi,\pi]\times[-\pi,\pi]$ using the same parameters values. Instead of assuming half-filling, the metric now includes only bands with energy below μ=0.50 at each $(k_x, k_y)$.

In Fig.5a, we have shown the contour plot of $g^{RR}_{n,\mu\nu}(k_x, k_y) = ReG^{RR}_{n,\mu\nu}(k_x, k_y)$ for the lowest band over $[-\pi,\pi]\times[-\pi,\pi]$, assuming half-filling, using the parameters values $t = 1$, $\varepsilon_A = 0.41$, $M_s = 0.95$, $\psi = \frac{\pi}{3}, \gamma = 0.23, J = 0.80, \Delta_{g0} = 0.50$, $\lambda_1 = 0.01, \lambda_2 = 0.001$, $\lambda_{rel} = 0.01$, $D_1 = 0.62$, and $D_2 = 0.53$. In Fig.5b, we have shown the contour plot of $g^{RR}_{n,\mu\nu}(k_x, k_y)$ summed over the occupied bands across $[-\pi,\pi]\times[-\pi,\pi]$, assuming half-filling, using the same parameters values. In Fig.5c, we have shown the refined contour plot of $g^{RR}_{n,\mu\nu}(k_x, k_y)$ summed over the occupied bands across $[-\pi,\pi]\times[-\pi,\pi]$ using the same parameters values. Instead of assuming half-filling, the metric now includes only bands with energy below μ=0.50 at each $(k_x, k_y)$. The k-grid is refined, revealing more detailed geometric features across the Brillouin zone. The plots capture how rapidly the quantum state changes with momentum. The plots show regions where the eigenstate varies rapidly with $k_x$, typically near avoided crossings, symmetry-enforced features, or flat-to-dispersive band transitions. Lines at $(k_x, k_y) \in \{0,\pi\}$ often reveal characteristic structure tied to the model's magnetic and orbital symmetries. In particular, Fig. c highlights collective geometric fluctuations of the filled space; large values can signal enhanced spread/localization tendencies in Wannier functions and sensitivity to perturbations.

We have obtained the Berry curvature contour plots (see Fig.6) of the occupied bands in Fig.2 separately using the Hamiltonian matrix for IAM. The numerical values of the parameters to be used in the plot are $t = 1$, $\varepsilon_A = 0.41$, $M_s = 0.95$, $\psi = \frac{\pi}{3}, \gamma = 0.23, J = 0.80$, μ = 0.50, $\Delta_{g0} = 0.50$, $\lambda_1 = 0.01, \lambda_2 = 0.001$, $\lambda_{rel} = 0.01$, $D_1 = 0.62$, and $D_2 = 0.53$. The Chern number of the occupied bands are computed on a refined k-mesh at chemical potential μ ≈ 0.5. With integer convergence within numerical tolerance $|C| < 10^{-5}$, we have found C=0.999994≈ +1. It must be stated that the popular practice is to follow the Fukui–Hatsugai–Suzuki (FHS)gauge-invariant lattice method [83] in order to accept only tolerances $< 10^{-8}$. We have not, however, used FHS method as numerical implementations may suffer from factors such as phase discontinuities in eigenvectors across the Brillouin zone, and branch cut issues when computing overlaps between neighboring k-points.

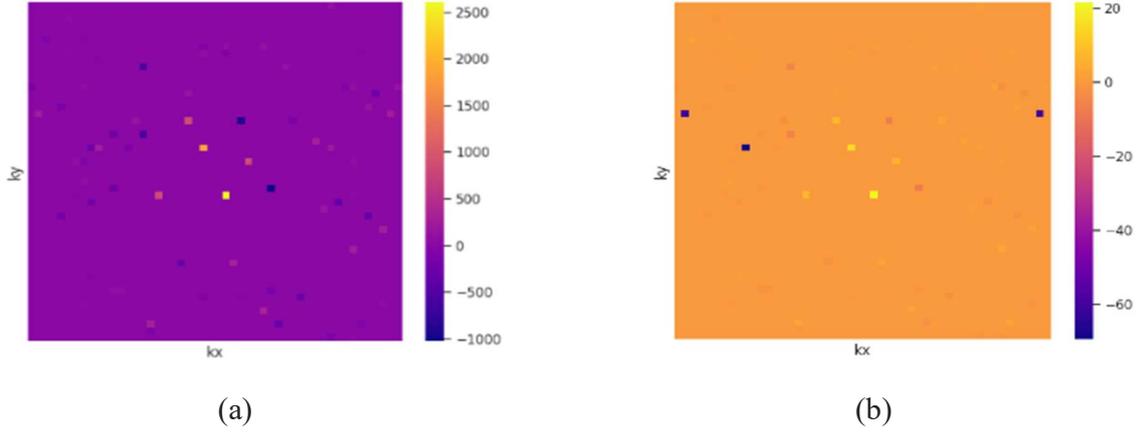

(a)                                            (b)

**Figure 6. (a) (b)** The contour plot of the Berry curvature of the occupied bands in Fig.2 using the Hamiltonian matrix for IAM. The numerical values of the parameters to be used in the plot are $t = 1$, $\varepsilon_A = 0.41$, $M_s = 0.95$, $\psi = \frac{\pi}{3}$, $\gamma = 0.23$, $J = 0.80$, $\mu = 0.50$, $\Delta_{g0} = 0.50$, $\lambda_1 = 0.01$, $\lambda_2 = 0.001$, $\lambda_{rel} = 0.01$, $D_1 = 0.62$, and $D_2 = 0.53$.

## 4. Future perspective and potential AM applications

A good heuristic[84] is that states of lower entropy are also less symmetric, in general. The symmetry $C_2$ has a lower order than $C_4$, so it is likely that the decrease in Boltzmann entropy is associated with the reduction in local symmetry from $C_4$ to $C_2$. We refer to our construction of a Hamiltonian for spinless fermions in two sublattices (A and B) on a square lattice, which yields a representative 2D model of insulating antiferromagnets. This model considers NN hopping, NNN hopping, an inversion symmetry-breaking term called the Semenoff-like mass, the on-site energy corresponding to the two sublattices, periodic magnetic flux (gauge field), and imagines each lattice site hosting a p orbital. This construction aims to reduce the local symmetries of the A and B sites from $C_4$ to $C_2$, preventing the lattice from reducing to a simple cubic structure with one atom in a unit cell. Thus, the Boltzmann entropy consideration is likely to predict a reduction in entropy (RIE). What insights does Shanon entropy consideration offer regarding RIE, given the model of insulating alternating magnets presented here? From a Shannon entropy perspective, symmetry breaking typically results in reduced entropy, as it introduces a preference for specific choices, thereby decreasing uncertainty within the model. This appears to align with the anticipated entropy reduction. The model's symmetry-breaking, coupled with the introduction of mass terms and gauge fields, should indeed yield lower entropy in the system, particularly in the insulating phase, consistent with expectations for a magnetically ordered or topologically gapped state. Exploring RIE within the context of the system under consideration emerges as a pivotal direction for future research.

A significant challenge for future AM applications and functionalization involves achieving controlled phase transition from conventional phases to the AM phase within a single material. Interestingly, research combining spin group symmetry analysis and ab-initio calculations demonstrated that ReO$_2$ undergoes a strain-induced AFM to AM phase transition under compressive strain [85], removing Kramer's degeneracy in the AM phase band structure within the non-relativistic regime, accompanied by a metal-insulator transition. These features illustrate the complexity and richness of AM systems, emphasizing their unique properties and potential applications in advanced materials science. The behavior of these materials exhibits a striking dissimilarity from conventional antiferromagnets. The distinctive characteristics of these AMs' give rise to numerous fascinating properties of fundamental significance and simultaneously open up avenues for potential technological applications in spintronics [86] and quantum information processing (QIP) [87]. The study of the authors in ref.[87] introduces a groundbreaking technique for entangling a mobile electron in an AM material with a confined electron in a quantum dot via spin-resolved scattering, achieving high fidelity with Bell's state and showcasing the promise of AMs for QIP.

## 5. Concluding remarks

The highlight of the paper is that the IAM system under consideration supports topological edge modes due to non-zero Chern number in a parameter window. In particular, an inversion symmetry-breaking term called the Semenoff-like mass ($M_s$) is assumed to be 0.95. For this value, the system hosts nonzero Chern number, indicating a topologically nontrivial phase. We have found that as $M_s$ increases, the band gaps close and reopen, causing the Berry curvature to redistribute and Chern numbers to vanish. This signals a topological phase transition from a Chern insulator to a trivial insulator.

Furthermore, it is essential to note that quantum anomalous Hall effect (QAHE) results from the sum of Berry curvatures of all occupied bands, whereas the anomalous Nernst effect (ANE) is calculated using the Berry curvature near the Fermi level. This distinction implies that an anomalous Hall effect of negligible/substantial significance does not necessarily imply ANE of negligible/consequential significance, due to differences in their underlying mechanisms [88.89]. The observations above emphasize the need to investigate the ANE to highlight the Berry curvature contribution.

## Appendix A

The Hamiltonian of AM insulator $H_{insul}(\mathbf{k})$ is given by the 4×4 matrix

$$\begin{pmatrix} A(\mathbf{k}) & T(\mathbf{k}) & 0 & \Delta(\mathbf{k}) \\ T^*(\mathbf{k}) & B(\mathbf{k}) & \Delta(\mathbf{k}) & 0 \\ 0 & \Delta(\mathbf{k}) & A(\mathbf{k}) & T(\mathbf{k}) \\ \Delta(\mathbf{k}) & 0 & T^*(\mathbf{k}) & B(\mathbf{k}) \end{pmatrix}, \quad (22)$$

where $A(\mathbf{k}) = a_1(\mathbf{k}) + ib(\mathbf{k})$, $B(\mathbf{k}) = a_2(\mathbf{k}) - ib(\mathbf{k})$, $T(\mathbf{k}) = t(\mathbf{k}) + it'(\mathbf{k})$, $\Delta(\mathbf{k}) = \left(\Delta_g(\mathbf{k}) + 2J(\mathbf{k})\right)$, $a_1(\mathbf{k}) = \varepsilon_A + M_s + \lambda_{rel} - 2(\lambda_1(\mathbf{k}) + \lambda_2(\mathbf{k}))$, $a_2(\mathbf{k}) = \varepsilon_B - M_s + \lambda_{rel} - 2(\lambda_1(\mathbf{k}) + \lambda_2(\mathbf{k}))$, $b(\mathbf{k}) = 2(D_1(\mathbf{k}) + D_2(\mathbf{k}))$, $\Delta_g(\mathbf{k}) = \Delta_{g0} \cos(4 \arctan(ak_y/ak_x))$, $\lambda_1(\mathbf{k}) = \lambda_1[\cos(ak_x + ak_y) + \cos(ak_x - ak_y)]$, $\lambda_2(\mathbf{k}) = \lambda_2[\cos(2ak_x) + \cos(2ak_y)]$, $D_1(\mathbf{k}) = D_1\hbar^2[\cos(ak_x + ak_y) + \cos(ak_x - ak_y)]$, and $D_2(\mathbf{k}) = D_2\hbar^2[\cos(2ak_x) + \cos(2ak_y)]$.

Here, in simpler form, the DM interaction could be written as $[D \sum_{\langle\langle I,j\rangle\rangle} \widehat{e_z} \cdot (S_i \times S_j)]$, where $D$ is the DMI constant, and $\widehat{e_z}$ is the unit vector along the z-axis as the system has out of plane spin-orbit coupling. The energy eigenvalues $E_n (n = 1,2,3,4)$ of $H_{insul}(\mathbf{k})$ are given by the equation $E_n^2 - E_n(A(\mathbf{k}) + B(\mathbf{k})) + A(\mathbf{k})B(\mathbf{k}) = |T(\mathbf{k})|^2 + \Delta^2(\mathbf{k}) \pm \Delta(\mathbf{k})(T(\mathbf{k}) + T^*(\mathbf{k}))$. It is easy to see that one of the conditions for $H_{insul}(\mathbf{k})$ being $PT$-symmetric is $\varepsilon_B = \varepsilon_A + 2M_s$. We prefer to work with $PT$-symmetric case for the sake of simplicity. The energy eigenvalues of $H_{insul}(\mathbf{k})$ are then given by

$$E_n(n = 1,2,3,4) = a_1(\mathbf{k}) \pm e_{\mp}(\mathbf{k}), \tag{23}$$

where $e_{\mp}(\mathbf{k}) = \sqrt{M_s^2 + (t(\mathbf{k}) \mp \Delta(\mathbf{k}))^{\wedge}2 + t'^2(\mathbf{k})) - b^2}$, $t(\mathbf{k}) = -2t \times (\cos ak_x + \cos ak_y) \cos(\psi)$, and $t'(\mathbf{k}) = -2t(\cos ak_x + \cos ak_y)\sin(\psi) + \gamma$. Thus, the second, more important, condition for $H_{insul}(\mathbf{k})$ being $PT$-symmetric is that the absolute magnitude of $2(D_1(\mathbf{k}) + D_2(\mathbf{k})) < \sqrt{M_s^2 + (t(\mathbf{k}) \mp \Delta(\mathbf{k}))^{\wedge}2 + t'^2(\mathbf{k}))}$ for all values of $\mathbf{k} = (k_x, k_y)$ in the first BZ. This puts a restriction on the choice of the DMI coefficients $(D_1, D_2)$. If the restriction is waived off, there will be possibility of EPs, at specific momenta, which represent branch-point singularities within non-Hermitian eigenvalue manifolds, characterized by the degenerate merging of eigenvalues and the coalescence of corresponding eigenvectors.

The right eigenstates linked to the eigenvalues of the Hamiltonian above could be written down in an explicit manner as

$$|u^{(n)}(k_x, k_y)\rangle = N_{n0}^{-\frac{1}{2}} \phi_n(k_x, k_y), \tag{24}$$

where $\phi_n(k_x, k_y)$ is the transpose of the row vector ($\psi_1^{(n)}(\mathbf{k})$  $\psi_2^{(n)}(\mathbf{k})$  $\psi_3^{(n)}(\mathbf{k})$  $\psi_4^{(n)}(\mathbf{k})$), $n = (1, 2,3,4)$, $\mathbf{k} = (k_x, k_y)$. The normalization factor $N_{n0}$ needs to be determined using the bi-orthonormality condition $\langle v^{(m)} | u^{(n)} \rangle = \delta_{mn}$. The right eigenstates

$$|u_n(k_x, k_y)\rangle = \frac{1}{N_{n0}^{\frac{1}{2}}(\mathbf{k})} \begin{pmatrix} \psi_1^{(n)}(k_x, k_y) \\ \psi_2^{(n)}(k_x, k_y) \\ \psi_3^{(n)}(k_x, k_y) \\ \psi_4^{(n)}(k_x, k_y) \end{pmatrix}, n = 1, 2, 3, 4, \tag{25}$$

The elements $\psi_j^{(n)}(\mathbf{k})$ (j=1,2,3,4) are given by

$$\psi_1^{(n)}(\mathbf{k}) = \wp_{10}^{(n)}(\mathbf{k}) + i\wp_{11}^{(n)}(\mathbf{k}), \psi_2^{(n)}(\mathbf{k}) = \wp_{20}^{(n)}(\mathbf{k}) + i\wp_{21}^{(n)}(\mathbf{k}), \psi_3^{(n)}(\mathbf{k}) = \wp_{30}^{(n)}(\mathbf{k}) + i\wp_{31}^{(n)}(\mathbf{k}), \psi_4^{(n)}(\mathbf{k}) = \wp_{40}^{(n)}(\mathbf{k}) + i\wp_{41}^{(n)}(\mathbf{k}), \tag{26}$$

where for the n$^{th}$ band

$$\wp_{10}^{(n)}(k) = \Delta(k)[E_n^2 - (a_1(k) + a_2(k))E_n + a_1(k)a_2(k) + b^2 + t^2(k) - t'^2(k) - \Delta^2(k)],$$

$$\wp_{11}^{(n)}(k) = 2\Delta(k)t(k)t'(k), \quad \wp_{20}^{(n)}(k) = 2\Delta(k)t(k)(E_n - a_1(k)), \quad \wp_{21}^{(n)}(k) = -2\Delta(k)t(k)b,$$

$$\wp_{30}^{(n)}(k) = t(k)[E_n^2 - (a_1(k) + a_2(k))E_n + a_1(k)a_2(k) + b^2 - (t^2(k) + t'^2(k)) + \Delta^2(k)],$$

$$\wp_{31}^{(n)}(k) = t'(k)[E_n^2 - (a_1(k) + a_2(k))E_n + a_1(k)a_2(k) + b^2 - (t^2(k) + t'^2(k)) - \Delta^2(k)],$$

$$\wp_{40}^{(n)}(k) = -(E_n - a_1(k))[E_n^2 - (a_1(k) + a_2(k))E_n + a_1(k)a_2(k) + b^2 + (t^2(k) + t'^2(k)) - \Delta^2(k)],$$

$$\wp_{41}^{(n)}(k) = b[E_n^2 - (a_1(k) + a_2(k))E_n + a_1(k)a_2(k) + b^2 + (t^2(k) + t'^2(k)) - \Delta^2(k)],$$

$$N_{n0}(k) = \wp_{10}^{(n)2}(k) + \wp_{11}^{(n)2}(k) + \wp_{20}^{(n)2}(k) + \wp_{21}^{(n)2}(k) + \wp_{30}^{(n)2}(k) + \wp_{31}^{(n)2}(k) + \wp_{40}^{(n)2}(k) + \wp_{41}^{(n)2}(k). \quad (27)$$

For the purpose of calculating Chern number (C), the z-component of RR-BC may now be written as **[88]**

$$\Omega_{xy}(k) = -2 \sum_{n \in Occupied\ Bands} Im \left\langle \frac{\partial u^{(n)}(k)}{\partial k_x} \middle| \frac{\partial u^{(n)}(k)}{\partial k_y} \right\rangle. \quad (28)$$

We use this formula to present the outline of the calculation of BC below. It is not difficult to see that for the present problem the product

$$\left\langle \frac{\partial u^{(\alpha)}(k)}{\partial k_x} \middle| \frac{\partial u^{(\alpha)}(k)}{\partial k_y} \right\rangle = \sum_{j=1,2,3,4} \left[ \left( P_{jx}^{(n)} P_{jy}^{(n)} + Q_{jx}^{(n)} Q_{jy}^{(n)} \right) + i \left( P_{jx}^{(n)} Q_{jy}^{(n)} - Q_{jx}^{(n)} P_{jy}^{(n)} \right) \right], \quad (29)$$

where

$$P_{jx}^{(n)} = -\left(\frac{1}{2}\right) N_{n0}^{-\frac{3}{2}} (\partial_x N_{n0}) \wp_{j0}^{(n)} + N_{n0}^{-\frac{1}{2}} (\partial_x \wp_{j0}^{(n)}),$$

$$Q_{jy}^{(n)} = -\left(\frac{1}{2}\right) N_{n0}^{-\frac{3}{2}} (\partial_y N_{n0}) \wp_{j1}^{(n)} + N_{n0}^{-\frac{1}{2}} (\partial_y \wp_{j1}^{(n)}), \quad (30)$$

$$Q_{jx}^{(n)} = -\left(\frac{1}{2}\right) N_{n0}^{-\frac{3}{2}} (\partial_x N_{n0}) \wp_{j1}^{(n)} + N_{n0}^{-\frac{1}{2}} (\partial_x \wp_{j1}^{(n)}),$$

$$P_{jy}^{(n)} = -\left(\frac{1}{2}\right) N_{n0}^{-\frac{3}{2}} (\partial_y N_{no}) \wp_{j0}^{(n)} + N_{n0}^{-\frac{1}{2}} (\partial_x \wp_{j0}^{(n)}), \quad (31)$$

$$(\partial_{\frac{x}{y}} N_{n0}) = 2 \sum_{j=1,2,3,4} [\wp_{j0}^{(n)} \left( \partial_{\frac{x}{y}} \wp_{j0}^{(n)} \right) + \wp_{j1}^{(n)} ( \partial_{\frac{x}{y}} \wp_{j1}^{(n)} )]. \quad (32)$$

The symbol $\partial_x$ ($\partial_y$) above stands for the differential coefficient $\frac{\partial}{\partial k_x}$ ($\frac{\partial}{\partial k_y}$). Now that we have calculated a formal expression for the BC of $n^{th}$ band, we still need to calculate various derivatives in (12). The expression of the anomalous Hall conductance (AHC) is given in refs.**[88,89]**.

The matrices $H_d(\mathbf{k})$ and $H_g(\mathbf{k})$, including the corresponding energy eigenvalues and the eigenvectors are given below:

$H_d(\mathbf{k}) =$

$$\begin{pmatrix} M - 2(\lambda_1(\mathbf{k}) + \lambda_2(\mathbf{k})) & \lambda_R(\mathbf{k}) & 0 & -2t(\cos ak_x + \cos ak_y) - \mu + \Delta_d(\mathbf{k}) \\ \lambda_R^*(\mathbf{k}) & -M - 2(\lambda_1(\mathbf{k}) + \lambda_2(\mathbf{k})) & -2t(\cos ak_x + \cos ak_y) - \mu + \Delta_d(\mathbf{k}) & 0 \\ 0 & -2t(\cos ak_x + \cos ak_y) - \mu + \Delta_d(\mathbf{k}) & -M - 2(\lambda_1(\mathbf{k}) + \lambda_2(\mathbf{k})) & \lambda_R(\mathbf{k}) \\ -2t(\cos ak_x + \cos ak_y) - \mu + \Delta_d(\mathbf{k}) & 0 & \lambda_R^*(\mathbf{k}) & M - 2(\lambda_1(\mathbf{k}) + \lambda_2(\mathbf{k})) \end{pmatrix},$$

(33)

and $H_g(\mathbf{k}) =$

$$\begin{pmatrix} M - 2(\lambda_1(\mathbf{k}) + \lambda_2(\mathbf{k})) + \Delta_g(\mathbf{k}) & \lambda_R(\mathbf{k}) & 0 & -2t(\cos ak_x + \cos ak_y) - \mu \\ \lambda_R^*(\mathbf{k}) & -M - 2(\lambda_1(\mathbf{k}) + \lambda_2(\mathbf{k})) + \Delta_g(\mathbf{k}) & -2t(\cos ak_x + \cos ak_y) - \mu & 0 \\ 0 & -2t(\cos ak_x + \cos ak_y) - \mu & -M - 2(\lambda_1(\mathbf{k}) + \lambda_2(\mathbf{k})) + \Delta_g(\mathbf{k}) & \lambda_R(\mathbf{k}) \\ -2t(\cos ak_x + \cos ak_y) - \mu & 0 & \lambda_R^*(\mathbf{k}) & M - 2(\lambda_1(\mathbf{k}) + \lambda_2(\mathbf{k})) + \Delta_g(\mathbf{k}) \end{pmatrix}.$$

(34)

The corresponding energy eigenvalues are

$$E_{d,n}(\mathbf{k}) = \pm\sqrt{M_d^2 + t_d^2 + |\lambda_R(\mathbf{k})|^2 \pm \sqrt{M_d^4 + 4t_d^2|\lambda_R(\mathbf{k})|^2 + 4M_d^2 t_d^2}}\ , \qquad (35)$$

$$E_{g,n}(\mathbf{k}) = \pm\sqrt{M_g^2 + t_g^2 + |\lambda_R(\mathbf{k})|^2 \pm \sqrt{M_g^4 + 4t_g^2|\lambda_R(\mathbf{k})|^2 + 4M_g^2 t_g^2}}\ , \qquad (36)$$

where

$$\lambda_R(\mathbf{k}) = \lambda_R(\sin ak_y + i \sin ak_x), \qquad (37)$$

$$t_d = -2t(\cos ak_x + \cos ak_y) - \mu + \Delta_d(\mathbf{k}), \qquad (38)$$

$$t_g = -2t(\cos ak_x + \cos ak_y) - \mu, \qquad (39)$$

$$M_d = M - 2(\lambda_1(\mathbf{k}) + \lambda_2(\mathbf{k})), \qquad (40)$$

$$M_g = M - 2(\lambda_1(\mathbf{k}) + \lambda_2(\mathbf{k})) + \Delta_g(\mathbf{k}). \qquad (41)$$